\documentclass[12pt,preprint]{aastex}

\shorttitle{New H$_2$ Jets in Mon R2}
\shortauthors{Hodapp}

\begin{document}

\title{NEW H$_2$ JETS IN MONOCEROS~R2}
\author{Klaus W. Hodapp\altaffilmark{1}} 

\altaffiltext{1}{
Institute for Astronomy, University of Hawaii,\\
640 N. Aohoku Place, Hilo, HI 96720,
\\email: hodapp@ifa.hawaii.edu }

\begin{abstract} 

We are presenting a wide-field image 
of the Mon~R2 star forming region
obtained with WFCAM on UKIRT
in the 2.12~$\mu$m filter centered on the H$_2$ 1--0 S(1)
emission line. We report the discovery of 15 new H$_2$ jets in
Mon~R2 and two in L~1646 and confirm most of these discoveries using archival 
Spitzer IRAC 4.5~$\mu$m and 8.0~$\mu$m images. We find that many of these protostellar
jets are found in projection against the outflow cavities of the
huge CO outflow in Mon~R2, suggesting that the jets may
be associated with an episode of star formation in Mon~R2 triggered
by this large, but now fossil, outflow.
We also study the spatial distribution of small, localized reflection nebulae
and find that these are distributed in the same way as photometrically identified
Class I sources.

\end{abstract}


\section{INTRODUCTION}

The region studied here was first identified by \citet{vdb66}
as an association of optical reflection nebulae in their list of such
objects and was labeled Monoceros~R2, in short Mon~R2.
The Mon~R2 molecular cloud complex extends over 3$\degr$ $\times$ 6$\degr$,
and the overall mass of this large molecular cloud has been estimated
by \citet{mad86} to be 9~$\times$~10$^4$ M$_\odot$.
The Mon~R2 molecular cloud contains several sites of star formation:
GGD~11, GGD~12-15 \citep{gyu78}, L~1646 \citep{car00}, 
and the Mon~R2 core, the main star formation site
that is the subject of this study.
The Mon~R2 star forming region lies between the two optical reflection
nebulae vdB~67 and vdB~69 \citep{vdb66} and has attracted considerable attention
as a site of massive star formation, as a young embedded cluster, and
as the source of one of the most powerful CO outflows known.

Even though Mon~R2 lies close to Orion in projection on the sky,
its distance
of 830$\pm$50~pc based on photometric parallaxes \citep{her76}
is almost double that of the Trapezium cluster, e.g. \citep{sta02}.
Despite its larger distance,
it is still close enough for detailed studies of the embedded
young objects. 
The mass of molecular gas in the Mon~R2 core has been determined by \cite{rid03} using 
$^{13}$CO and C$^{18}$O maps. 
The rarer C$^{18}$O traces denser gas and resulted
in a gas mass of 1826~M$_\odot$,
while the more abundant $^{13}$CO also
samples the tenuous outer regions of the molecular core and gave a mass
of 2550~M$_\odot$. 
The molecular gas distribution is single-peaked
with a slight indication of extended filaments to the south and south-west. 
To put this region into perspective, 
Mon~R2 is smaller than the Orion Trapezium region in spatial 
extent, molecular mass, and number and total mass of embedded stars. It is, 
however, not far behind in many of these parameters.

Mon~R2 has been noted as a source of a bipolar CO outflow by \citet{lor81}. 
This main Mon~R2 molecular outflow was found by \citet{wol90}
to be one of the dynamically oldest (1.5$\times10^5$~y), 
largest (7 pc), and most massive (100 $M_{\odot}$) molecular outflows. 
Based on CO maps with higher spatial resolution than those of
\citet{wol90}, \citet{mey91} argued that the Mon~R2 CO outflow must
consist of two bipolar outflows: one component seen almost head-on to explain the
overlapping red and blue shifted velocity components seen in projection
against the central cluster, and one component more
inclined against the line of sight, to explain the bipolar outflow features.
Both components could be modeled as paraboloidal shells with a
linear velocity field that could result from velocity-sorting of
clumpy outflow material.
The detailed study of CO emission and multiple transitions of CS by \citet{taf97} 
showed that the
massive outflow has created an hourglass-shaped cavity in the molecular cloud 
centered on the cluster
of massive young stars. The cavity walls coincide with the limb-brightened shells of
the blue-shifted outflow lobe. The axis of the paraboloidal outflow
shell and cavity are
oriented roughly at a position angle of -30$^\circ$.

The central group of luminous stars and the associated nebulosity
were discovered in the near infrared by \citet{bec76}.
Much of the attention has been focused on the ring-like reflection
nebula around IRS~1 discovered by \citet{bec76} and studied in more detail by \citet{hod87},
on the details of the cluster's most luminous individual source (IRS~3) summarized in the most
recent work by \citet{pre02}, and on the
distribution of molecular matter and the large outflow, summarized in \citet{taf97}. 
In embedded young clusters like Mon~R2, near-infrared (J, H, and K) photometry
is a very useful tool for identifying young cluster members on the basis of their
color excess over the reddened photospheres of old background stars. The first
such study of Mon~R2 was done by \citet{car97}.
Most recently, the Initial Mass Function (IMF) in the Mon~R2 cluster core was studied in detail
by \citet{and06} on the basis of HST/NICMOS data. They find
that the ratio of the number of stars below solar mass to the number of objects in the brown
dwarf mass range is similar to that found in the Trapezium cluster and in
IC~348.

Compared to the widely used J, H, and K images used to characterize the stellar
population of young clusters, 
the Spitzer IRAC images in four bands ranging from 3.6~$\mu$m
to 8.0~$\mu$m allow the identification of younger and more deeply embedded objects.
Further, they also allow the identification of objects with purely longer-wavelength infrared excess, 
indicating passive reprocessing
disks, e. g. \citep{gut05} and \citep{kum07}. 
On the basis of near-infrared J, H, and K photometry, and of Spitzer mid-infrared photometry,
\citet{gut05} found that Class I objects in Mon~R2 are distributed in a filamentary pattern,
while Class II objects are distributed more widely and evenly.
However, even these IRAC bands are not sampling the youngest 
objects in a cluster completely,
in particular those Class 0 and I objects that are oriented with their disks seen edge-on. 
Fortunately, those objects can still be found indirectly. 
The substantial accretion activity in the Class 0 and I phases of
low-mass star formation is invariably associated with outflow activity, as was first
noted by \citet{and93}. These outflows
easily break out of the dense molecular disk and the envelop enshrouding the young star,
and are therefore readily detected in the near-infrared, due to the shock-excited emission
at the interface between the outflow and the ambient molecular material, and in
shocks within the outflow. The most easily accessible near-infrared shock-excited
emission lines are the [FeII] line at 1.644~$\mu$m and the H$_2$~1--0~S(1) line at
2.122~$\mu$m, the line studied here.
Of these two emission lines, the H$_2$ 1--0 S(1) line traces lower excitation levels and is excited
in virtually all deeply embedded jets propagating in molecular clouds. 
Therefore, shock-excited emission of molecular hydrogen is the earliest near-infrared signpost of
outflow activity in a very young object and is usually observable long before
the central star becomes observable in the near or mid infrared.
Molecular hydrogen jets tend to have morphological features that are distinct from
other forms of H$_2$ emission in star forming regions, and are therefore easily
recognized in imaging surveys.

Powerful jet activity is a byproduct of the short-lived main accretion phase of a forming
star. When the accretion and jet activity fades after about 10$^5$ y, 
the envelope of molecular gas around a young star
usually has formed two outflow cavities. 
When properly oriented, scattered light from these cavity walls 
gives the object the appearance
of a bipolar nebula, or a cometary nebula if only one cavity can be seen. Objects of this
morphology overlap with the Class I objects that can be identified photometrically using the
Spitzer IRAC bands.
Therefore, imaging
surveys of star forming regions for protostellar jets and bipolar or cometary nebulae
are a powerful tool for
discovering very young (Class 0 or I) objects in or soon after their main
accretion phase. 
Thanks to the recent availability of large-format near-infrared cameras such surveys
can now be conducted efficiently.

In this paper, we report the discovery of 17 extremely young objects
in and around the Mon~R2 star forming region by means of imaging their
collimated outflows in the H$_2$ 1--0 S(1) emission line at 2.12~$\mu$m.
By comparison of the 2.12~$\mu$m image and the Spitzer IRAC 4.5~$\mu$m 
and 8.0~$\mu$m images we discuss the most plausible identification of the source of each outflow.
The large number of outflows indicates that star formation is active all
through the Mon~R2 molecular cloud. We will discuss the spatial distribution of the newly
discovered extremely young stars in relation to the large, old molecular outflow in Mon~R2.

\section{OBSERVATIONS AND DATA REDUCTION}

\subsection{UKIRT WFCAM 2.12 $\mu$m Imaging}

The purpose of this project was not the detailed study of any particular
jet, but rather the discovery of new objects in the Mon~R2 star forming 
region that indicate the presence of
stars in their youngest phase of evolution. 
Also, completeness in the discovery of H$_2$ outflows was not a realistic
goal for this survey of a region with such extensive reflection nebulosity
and, most likely, also fluorescently excited H$_2$ emission (e. g., \citet{bla76})
that both tend to confuse any search for H$_2$ outflows.
Therefore, in the interest of maximizing
the scientific return of the observing time used, we did not obtain an image
at a continuum wavelength adjacent to the S(1) line, or in a broad K-band
filter to distinguish between emission line features and continuum reflection
nebulosity. As discussed below, the identification of H$_2$ jets relied
on morphological arguments and confirmation, if possible, by archival Spitzer IRAC images.

Wide-field near-infrared images of the Mon R2 region
were obtained in the night of October 20, 2005, UT, at the
United Kingdom Infrared Telescope (UKIRT). The 
Wide-Field Camera (WFCAM) \citep{cas07} was used with a filter 
centered
at 2.121 $\mu$m with $\Delta \lambda$ = 0.021 $\mu$m
that includes the H$_2$ 1--0 S(1) line.
The field was roughly centered on
the well-studied Mon R2 cluster of infrared sources.
The WFCAM uses four Teledyne (formerly Rockwell) HAWAII-2 2048$\times$2048 HgCdTe infrared detector arrays
at a pixel scale of 0.4$\arcsec$ pixel$^{-1}$,
arranged in the camera focal plane with a spacing of 0.94 of the
detector size between the devices. To achieve full coverage of a 0.75 square degrees
region of the sky, four telescope pointings are required, which
in WFCAM terminology are called a ``tile''.
At each tile position, a five-point dither pattern (3.2$\arcsec$ spacing) is used to
avoid bad pixels and other detector array artifacts. Finally, at each
dither position, a four-point micro-step pattern with a half-pixel (0.2$\arcsec$)
step size is used to oversample the image.
For our observations, an on-chip exposure time of 40~s was chosen, and the detectors
were read out in non-destructive mode, sampling the accumulating signal
41 times during the integration time. The readnoise achieved in this mode
of operation is $\approx$~20~e$^-$ rms. 
The on-sky integration time was 800~s,
except in the overlap regions of the tile. 
With a distance to Mon~R2 of 830~pc, one arcminute in our images corresponds to a projected
distance of 0.24~pc.

The data were processed by the Cambridge Astronomical Survey
Unit (CASU) using the procedures described by \citet{dye06}.
Astrometric calibration relative to the 2MASS catalog \citep{skr06} was 
performed and a plate solution was written into the file headers.
All coordinates in this paper therefore are given in J2000.0 
and are indirectly based on the 2MASS coordinate system.
After retrieval of the reduced data from the WFCAM Science 
Archive the final assembly of the full tile image was done with
IRAF scripts that used the image header information
to assemble the individual images into a tangential projection
image of the full tile.

The central region of the resulting image is shown in Fig.~1.
The image shows the well-known Mon~R2 cluster near the center, and the
two reflection nebulae to the west (vdB~67) and east (vdB~69) of it. Filamentary 
emission extends mostly to the north and east of the main star
forming region. The lowest contours of the blue and red shifted CO maps of \citet{wol90} are 
superposed on Fig.~1. There is no large-scale feature visible that would
be morphologically associated with the main CO outflow 
centered on IRS~2 (identified in Fig.~2) that 
has a rather poorly defined outflow axis at P.A. $\approx$-45$^\circ$ \citep{wol90}
or with the outflow cavity whose axis is oriented at $\approx$-30$^\circ$
\citep{taf97} and \citep{cho00}.

\subsection{Archival Spitzer Images}

The IRAC camera is the main near and mid-infrared imaging instrument of
the Spitzer cryogenically cooled space telescope \citep{faz04}.
All the four IRAC channels contain emission lines of H$_2$. 
Of the four IRAC channels, band~2 centered at 4.5 $\mu$m is particularly
well suited to detect H$_2$ emission. From a discussion of different shock models, \citet{smi05} concluded
that shock-excited H$_2$ emission in
the 4.5 $\mu$m band is an order of magnitude brighter than in the other channels
for a wide range of shock conditions.
The IRAC band 2 (4.5 $\mu$m) contains the H$_2$ 0-0 S(9) line at 4.694 $\mu$m and
emission in this line shows morphological features closely identical to those
seen in H$_2$ 1--0 S(1) at 2.122 $\mu$m, but is less affected by dust extinction. 
Therefore, despite the fact that the
broad IRAC band 2 also contains other important molecular features such as CO ($\nu$=1--0
at 4.45 - 4.95 $\mu$m) and the atomic Hydrogen Br$\alpha$ line at 4.052 $\mu$m, images
in this band often show very similar jet morphology to 2.12~$\mu$m S(1) images
\citep{smi06}.
By contrast, \citet{smi05} find that IRAC band 1 (3.6~$\mu$m)
is dominated by vibrationally excited lines with higher excitation energy than
the 1--0 S(1) line, so that shock features appear sharper in band 1 than in a
2.12$\mu$m 1--0 S(1) image.
The Spitzer 3.6 $\mu$m, 5.8 $\mu$m, and 8.0 $\mu$m bands also contain bright PAH
emission at 3.3 $\mu$m, 6.2 $\mu$m, and 7.7 $\mu$m, respectively. In a region
like Mon~R2, this strong
and often filamentary emission tends to mask shock-excited emission from jets.  

Spitzer IRAC images were downloaded from the Spitzer archive.
The data used here were originally obtained for a program by
G. Fazio. Partial results of their study of the distribution
of stars based on JHK and Spitzer/IRAC colors have been presented
by \citet{gut05}.

The available archival Spitzer data do not cover the whole field of our WFCAM
observations, so that we do not have the corresponding longer-wavelength data
for all of the outflows discovered here. For those objects where all Spitzer 
bands are available, we
discuss the 4.5~$\mu$m and 8.0~$\mu$m data. The 4.5~$\mu$m data usually show the 
outflows, but at relatively low signal-to-noise ratio and with poorer spatial
resolution that the UKIRT data. The 8.0~$\mu$m data do not
usually show the outflow itself, but in many cases show the molecular clump
around the driving source either in thermal dust emission or in some cases
in absorption against PAH background emission. The 8.0~$\mu$m images
therefore primarily serve to confirm the identification of the driving source.
In some cases the band 2 (4.5~$\mu$m) data were not available and for one object
band~1 data
(3.6 $\mu$m) data are shown instead for the purpose of source confirmation.

\subsection{Criteria for Identifying H$_2$ Jets}

This simple imaging project was motivated by the success of similar observations
in identifying H$_2$ jets,
e.~g. the early survey of NGC~1333 by \citet{hod95}, the extensive survey
of Orion~A by \citet{sta02}, the survey of the distant massive star forming region
W~51 by \citet{hod02},
and the recent work on DR21/W75 by \citet{dav07}.
Well over one hundred of these collimated H$_2$ jets have now been imaged and
their morphology is quite unique and well recognizable. 
Empirically, shock excited H$_2$ emission can, in most cases, be distinguished from
fluorescently excited H$_2$ emission on the basis of morphological arguments.
The physical basis for this is that the cooling time of shock-excited molecular
hydrogen is typically only a few years \citep{shu78}, during which time shock fronts travel
only a fraction of an arcsecond in typical star forming regions. Shock-excited
emission therefore usually appears as small but resolved knots, filaments,
and bow shocks. In contrast, fluorescently excited molecular hydrogen emission
varies on the scale of typical density inhomogeneities in molecular clouds and
therefore tends to be much more smoothly distributed. The same is true for
large-scale continuum reflection nebulosity. 

Our criteria for identifying H$_2$ jets in Mon~R2 therefore were that
they must consist
of a string of resolved H$_2$ emission knots. 
If a bow shock is found at the end of a string of emission knots, this
is regarded as additional and convincing evidence for the existence 
of a jet. In cases where both sides (lobes) of a jet are visible, a high
degree of central symmetry in the distribution of emission knots serves
as supporting evidence for the identification of a jet.
Sometimes, but not always, faint filamentary emission 
arising from the outflow cavity walls connects
the individual emission knots. 
If such filamentary emission is seen, it serves as additional evidence
supporting the identification of a jet.
The H$_2$ jets identified by these criteria are listed in Table~1, and identified
in Fig.~1 as large circles. 

Finally, many young stars show reflection nebulae of
bipolar or cometary morphology or nebulae otherwise 
clearly associated with individual stars, properties
that we summarize by the term ''localized''. 
In addition to their characteristic morphology, they show
a more pronounced drop in surface brightness
away from the illuminating central star than the shock emission knots
in H$_2$ jets.
In Table~2, we list such localized nebulae
that were found on our 2.12$\mu$m image on Mon~R2, but
were not classified as likely H$_2$ jets. 
Nebulous features that are simply knots 
in the large-scale nebulosity permeating the center of the Mon~R2 cluster are not
listed in Table~2. 
Similar to the situation with the H$_2$ jets,
this list cannot be expected to be complete, due to sensitivity limits, high extinction
in some parts of the molecular cloud, and confusion with the large-scale reflection 
nebulosity and filamentary emission present in the Mon~R2 region. The localized reflection nebulae
are indicated as small squares in Figs.~1 and 2.

The K' survey of known CO outflow sources by \citet{hod94}
showed that many of these outflow sources are associated with such localized reflection nebulae.
Reflection nebulae, in particular of bipolar or cometary morphology, are produced by
light scattered from the inside walls of the cavity produced 
by an outflow in the molecular material surrounding a young star.
While the
association of reflection nebulae with young age of an individual star and with outflow
activity is less direct 
than for H$_2$ jets, the presence of large numbers of these localized reflection nebulae
nevertheless indicates a population of very young stars, probably stars just after their main
accretion and outflow phase.

\section{RESULTS AND DISCUSSION}

\subsection{The Environment of the Mon~R2 Cluster}

Fig.~1 shows an extended system of emission filaments to the north and east of the cluster, 
besides the known reflection nebulae vdB 67 and 69, the nebulosity
associated with the Mon~R2 cluster, and the individual H$_2$ jets 
reported here.

These filaments coincide with the optical reflection nebula
vdB~68, and are also strongly visible in all Spitzer IRAC bands,
probably indicating a combination of H$_2$ and PAH emission
fluorescently excited by the
illuminating star of vdB~68, a B2V star \citep{her76}.
The filaments seen in Fig.~1 are oriented at a position angle of $\approx$ 35$\degr$ ,
similar to some of the filamentary structure seen in the optical
reflection nebula. It should be noted that this orientation of the
filaments is similar to the polarization angle of background star polarization
in the region of vdB~68 as measured by \citet{jar94} in the R band, which
indicates the projected direction of the local magnetic field.

On the Digital Sky Survey (DSS) blue and red
images, an extensive system of filaments is seen in absorption against the
background reflection nebulosity in the area south and east of vdB~67,
the Mon~R2 cluster, and vdB~69. These filaments are also oriented along
the direction of the local magnetic field, as measured by the R-band polarimetry
of \citet{jar94}. Our 2.12~$\mu$m S(1) image does not show any emission from these filaments 
located in the south east of the embedded cluster.  

From the fact that these systems of filaments are readily visible at optical
wavelengths, we conclude that they lie on the front side of the Mon~R2 molecular
core. Since the polarization vectors of more deeply embedded sources in and near
the Mon~R2 cluster \citep{hod87} follow the pattern of the magnetic field measured to the south
and west of the cluster, and since the cluster is located at the intersection 
of the two patterns of magnetic field direction, it can be speculated that the 
formation of the Mon~R2 cluster was triggered by a collision to two molecular
clouds, as discussed by \citet{jar94}.

The core of the Mon~R2 cluster and the bright reflection nebulosity permeating it
are shown in more detail in Fig.~2.
From polarization measurements of this ring-like infrared reflection nebulosity
near the core of the Mon~R2 cluster by \citet{hod87}, it is clear that IRS~2 is its 
dominant source of illumination. The brighter source IRS~3 does
not significantly contribute to the illumination (as measured by polarization) of 
the ring-like nebula, and therefore must be located either in front or behind the
nebula and the core of the Mon~R2 cluster. 
Some elongated emission features are seen east of IRS~3 and lie in a direction similar
to the filaments observed near vdB~69. This would suggest that IRS~3 is illuminating
some of these filaments and that it lies in front of the rest of the Mon~R2 cluster.

\subsection{Previously Known Outflows}

The overwhelming size and total mass of the main Mon~R2 CO outflow \citep{lor81}
makes it difficult to find evidence for other outflows in the
velocity field measured in $^{12}$CO(2-1) or $^{13}$CO(2-1). 
It should be noted that \citet{mey91} have argued that the distribution of
high velocity CO gas in this main outflow can only be explained
by a superposition of two independent outflows with different 
orientations, one almost head-on and the other more inclined to
produce the bipolar appearance. 
Despite the overwhelming emission from these two main outflows, 
two additional possible outflows associated
with the Mon~R2 cluster have been discovered by
CO emission line mapping by \citet{taf97} and a third
possible outflow (a microjet from IRS~3) was found by \citet{pre02}.

The $^{12}$CO(2-1) and $^{13}$CO(2-1) velocity channel maps 
of \citet{taf97} (their Figs. 3 and 4)
show a redshifted high velocity
component (Mon~R2-N) about 80$\arcsec$ to the north of the central cluster,
coinciding with ``Object 1'' of \citet{coh77}. This object
is the brightest of a group of embedded stars north of the ring-like
reflection nebula, and is itself embedded in reflection nebulosity
with a complex filamentary structure shown in Fig.~2. 
The brightest star in this sub-group of the Mon~R2 cluster is
located at 6$^h$7$^m$45$\stackrel{s}{.}$9, -6$\degr$21$\arcmin$47$\arcsec$. 
Also, \cite{taf97} discuss a strongly blueshifted $^{12}$CO feature
(-100$\arcsec$,-40$\arcsec$) from IRS~1 that also coincides with
an enhancement of CS emission, indicating the presence of a deeply
embedded object. We refer to this object by the initials of the
authors as ''TBWW97 CO-100,-40''. 
At its position (6$^h$7$^m$39$\stackrel{s}{.}$2, -6$\degr$23$\arcmin$44$\arcsec$), we note a faint,
slightly elongated infrared object on the S(1) image, surrounded by several small
patches of nebulosity (Fig.~2 and Fig.~3, upper panels). 

The Spitzer 4.5~$\mu$m image shows the same object morphology
and confirms that the marginally extended 2.12~$\mu$m object is not an embedded star.
The Spitzer 8.0~$\mu$m image shows extended emission in this region.
The high-velocity CO gas and the shock-excited H$_2$ emission
indicate a spatially compact outflow, without a clear jet morphology.
This object is shown near the western edge of Fig.~2 (labeled ''CO''). and in the first panel
of Fig.~3. 

The infrared source IRS~3 was noted by \citep{bec76} to be extended
and was later resolved into a number of sources. The highest spatial
resolution image are the speckle interferometry results by \citet{pre02}
that list point sources IRS~3 A - F, of which A, B, and C are associated with
strong reflection nebulosity, and B is possibly associated with a microjet
of about 0.5$\arcsec$ length.
The detailed CO velocity maps by \citet{gia97} show that IRS~3 is not associated
with the big Mon~R2 CO outflow, but that it is a separate source of
very high-velocity CO outflow, even though a clear bipolar feature was not
detected.

\subsection{The Spatial Distribution of H$_2$ Jets}

The newly discovered H$_2$ jets are overlayed on Fig.~1 as open circles with
labels. Also overlayed are the lowest contours of the blue and redshifted 
features in the CO map of \citet{wol90}. Their map was chosen over the
higher spatial resolution maps of \citet{mey91} and \citet{taf97} since
it covers a larger area with higher sensitivity to faint extended features.
The only sites of star formation lying clearly outside of the CO outflow
lobes are the newly discovered HOD07 1 to the west of the main cluster,
and the previously known young object GGD~11. The longest of the newly
discovered H$_2$ jets (HOD07 13) lies just south of the redshifted CO lobe.

The CO outflows probably have a shell structure with a relatively empty
cavity near the outflow axis \citep{mey91}. Since the outflowing material interacts
turbulently with the molecular material of the ambient cloud, triggered
star formation would be expected near the interface between the shell and the ambient cloud.
In projection, the highest density of triggered star formation sites
would be expected where we look tangentially
along the shell wall, and a smaller number where we look onto the front
and back sides of the outflow shells. 
The overall distribution of newly found H$_2$ jets outside of the central
cluster roughly matches this expectation. Note in particular that the
small cluster associated with HH~866 is projected against the tip of the blueshifted 
CO contour of \citet{wol90}, and that the group of H$_2$ jets HOD07 2, 3, 6, 8,
and 9 lie close to the eastern edge of the blueshifted CO contour, and behind
two secondary blue-shifted CO emission maxima.
These spatial coincidences strongly suggest that sites of recent star formation
outside the main Mon~R2 cluster may have been triggered by turbulent interaction
of the massive main Mon~R2 outflow with the ambient molecular material.

In the following, we discuss the newly discovered H$_2$ jets individually
on the basis of their morphology and, when available, by a comparison
of the near-infrared 2.12 $\mu$m S(1) image to longer wavelength Spitzer images.
The objects are numbered in the sequence of increasing R.A., and labeled by the
prefix HOD07. All coordinates are given in the J2000.0 system.
To allow an easy comparison of their sizes, the overview 2.12 $\mu$m images based
on the UKIRT/WFCAM data are all shown on the same spatial scale. Similarly, the detailed
views in the figures comparing WFCAM and Spitzer/IRAC results are all on the same, 
but finer, spatial scale, with the exception of Fig.~6, where this was not practical.

The small localized reflection nebulae that we found on our 2.12 $\mu$m image, 
but did not classify as jets,
are listed in Table~2 and are indicated by small squares in Figs. 1 and 2.
These localized reflection nebulae are concentrated to the south of the Mon~R2 cluster, closer to the
cluster than the H$_2$ jets found to the south and south west of the cluster.
Since the reflection nebulae represent a more advanced state of star formation 
than the jets, this distribution might indicate a sequence of triggered star
formation to the south of the main Mon~R2 cluster, in the region where the
blue and red shifted CO contours overlap.

\subsection{Discussion of Individual H$_2$ Jets}

\subsubsection{HOD07 1}
The H$_2$ jet HOD07~1 has a very symmetric bipolar appearance, as is shown in Fig.~4. 
We note a high degree of similarity in the shape and relative intensity of corresponding
knots in the two lobes of the jet.
We interpret the two extended patches
of nebulosity $\approx 3\arcsec$ east (labeled 1E) and west (labeled 1W) of 
the apparent center of symmetry at
coordinates 6$^h$7$^m$10$\stackrel{s}{.}$ -6$\degr$26$\arcmin$34$\arcsec$ 
(marked by a circle in Fig.~4) as emission or scattered
light from the two outflow cavities, and assume that the driving source is hidden 
from direct view and located near that position in the
dark region between those two emission knots. 
From this position, the eastern jet lobe has a position angle of $\approx$~60$\degr$.
Since HOD07~1 is not included in the publicly released Spitzer images of the Mon~R2 region,
this position of the central source could not be confirmed by longer wavelength data.

Going symmetrically outwards from
the assumed position of the driving source, there is a pair of faint emission knots at 
6$^h$7$^m$9$\stackrel{s}{.}$5, -6$\degr$26$\arcmin$44$\arcsec$ (2W)
and 6$^h$7$^m$11$\stackrel{s}{.}$6, -6$\degr$26$\arcmin$22$\arcsec$ (2E), respectively. 
The next pair of relatively faint emission knots is at
6$^h$7$^m$9$\stackrel{s}{.}$0, -6$\degr$26$\arcmin$46$\arcsec$ (3W) and 
6$^h$7$^m$12$\stackrel{s}{.}$4, -6$\degr$26$\arcmin$16$\arcsec$ (3E).
Further away from the central source we find two complex systems
of bright emission knots, containing two (west) and three (east) individual knots,
centered at 
6$^h$7$^m$8$\stackrel{s}{.}$2, -6$\degr$26$\arcmin$52$\arcsec$ (4W) and 
6$^h$7$^m$12.9, -6$\degr$26$\arcmin$11$\arcsec$ (4E), respectively.
Both of these emission regions show a bend of $\approx$ 45$^\circ$ to the left in the direction
of motion. 
To the west of the central source, at 6$^h$7$^m$7$\stackrel{s}{.}$5, -6$\degr$26$\arcmin$54$\arcsec$ (5W)
there is one additional very strong emission knot of approximately triangular shape with no
detectable counterpart in the east.
Finally, the most distant, approximately symmetric set of faint shock emission features is
located at 6$^h$7$^m$6$\stackrel{s}{.}$3, -6$\degr$26$\arcmin$57$\arcsec$ (6W) and 
6$^h$7$^m$16$\stackrel{s}{.}$7, -6$\degr$25$\arcmin$53$\arcsec$ (6E)
On the eastern side, this
faint extended nebulosity coincides with the images of several stars, 
but there is no evidence to suggest
that these stars are physically associated with the outflow.

It is noteworthy that the bend in the otherwise straight bipolar jet coincides with the
strongest emission knots. Bent or S-shaped jets as a result of disk precession in a binary
system were discussed by \citet{ter99} and \citet{bat00}. 
We speculate that the main accretion and outflow event that produced
these strong systems of emission knots was triggered by a close stellar encounter, most likely with
a binary star component on an elliptical orbit, and that this encounter was also responsible 
for a change in the disk orientation and therefore the jet direction. 

\subsubsection{HOD07 2}
The extended S(1) emission with shock-like morphology labeled HOD07 2
in Fig.~5 at 6$^h$7$^m$30$\stackrel{s}{.}$5, -6$\degr$11$\arcmin$51$\arcsec$
consists of two distinct knots of emission, suggesting that this is
a short jet (P.A. $\approx$~120$\degr$) originating from a star that itself is too obscured to
be directly visible. The position of the more compact shock emission knot
is the position given above and in Table~1. 
The Spitzer 8 $\mu$m image (Fig.~6, bottom panel)
shows extended emission centered on a point source 
at 6$^h$7$^m$30$\stackrel{s}{.}$9, -6$\degr$11$\arcmin$48$\arcsec$.
While this point source indicates other ongoing star formation
activity in this region, its position does not coincide with the jet HOD07~2 and it therefore
cannot be identified with the driving
source of this jet.

\subsubsection{HOD07~3}
The bow shock HOD07~3 (Fig.~5) at 6$^h$7$^m$31$\stackrel{s}{.}$6, -6$\degr$12$\arcmin$47$\arcsec$ 
is probably driven by a source some distance away
to the south east of the shock. 
Projecting back from the bow shock in this direction (P.A. $\approx$~147$\degr$)
we find a string of faint, slightly extended
knots of 2.12~$\mu$m emission.
These faint emission features, indicated by arrows in Fig.~5,
are located at 6$^h$7$^m$33$\stackrel{s}{.}$2, -6$\degr$13$\arcmin$03$\arcsec$, at 
6$^h$7$^m$34$\stackrel{s}{.}$1, -6$\degr$13$\arcmin$24$\arcsec$, 
and at 6$^h$7$^m$39$\stackrel{s}{.}$3, -6$\degr$14$\arcmin$01$\arcsec$.
Overall, they appear to form a jet with multiple, faint
internal shocks, ending in the bright bow shock.
All these emission knots are confirmed by the Spitzer 4.5~$\mu$m data,
even though only the knot closest to the bow shock is included
in Fig.~6, where we show the bow shock at 2.12~$\mu$m, 4.5~$\mu$m, and 8.0~$\mu$m.
However, the position of the driving source of this jet could not
be identified, neither on the WFCAM image nor the Spitzer images.
We speculate that the jet emerges from deep within the molecular
cloud and that only the terminating bow shock moving towards the
observer is seen at relatively low extinction.
Therefore, as indicated in Fig.~5, the position listed for this jet in Table~1 is that of the 
bow shock. 

\subsubsection{HOD07 4}
Two knots of H$_2$ emission are visible to
the west of the star 
at 6$^h$7$^m$35$\stackrel{s}{.}$3, -6$\degr$20$\arcmin$00$\arcsec$ 
in Fig.~7 and the top panel of Fig.~8. 
However, the extended nebulosity associated with the south-western of these knots
is not connected with this star, strongly suggesting that this star may not be physically associated 
with the H$_2$ emission. 
A second, fainter system of H$_2$ emission with a bow shock morphology is visible
at 6$^h$7$^m$35$\stackrel{s}{.}$9, -6$\degr$19$\arcmin$56$\arcsec$, 
to the east-northeast of this star.
The IRAC 3.6 $\mu$m (not shown here) 
and 4.5 $\mu$m images (Fig.~8, top center) show faint curved nebulosity connecting
the 2.12~$\mu$m features with a very faint star 
at position 6$^h$7$^m$37$\stackrel{s}{.}$6, -6$\degr$19$\arcmin$54$\arcsec$ that is visible on the
2.12~$\mu$m UKIRT image as well as on the two shorter wavelength IRAC images. 
There is no evidence for a counterjet, bent or straight,
emerging from the position of that faint star. 
Unfortunately, the longer wavelength IRAC archival images
at this position are not of useable quality. 
From the available data we conclude that the S(1) shock emission
and the S-shaped feature in the IRAC 3.6 $\mu$m and 4.5 $\mu$m images represent the less obscured parts
of a jet emerging from the star at 6$^h$7$^m$37$\stackrel{s}{.}$6, -6$\degr$19$\arcmin$54$\arcsec$. 
To explain the bending of the jet, it must be emitted from a precessing disk, similar to the case of
IRAS 03256+3055 in NGC~1333 discussed by \citet{hod05}.

\subsubsection{HOD07 5}
The position given here for HOD07~5: 
6$^h$7$^m$38$\stackrel{s}{.}$7, -6$\degr$21$\arcmin$14$\arcsec$, indicated by a circle in
Fig.~7 and Fig.~8, middle panel,
is that of a faint extended emission 
feature near the apparent symmetry center of
a system of shocks oriented at a position angle of
$\approx$~32$\degr$. 
This central object is visible on the UKIRT 2.12~$\mu$m image,
but is much brighter on the Spitzer 4.5~$\mu$m image, indicating heavy obscuration.
Also, the 2.12~$\mu$m image of this central feature shows a marginally detected bifurcation, 
indicating a disk seen
in absorption against the extended emission or a disk shadow.
We tentatively conclude that the
driving source is not directly visible at near-infrared wavelengths, but that emission 
and/or scattered light very
close to the source has been detected, and that the collimating disk is seen nearly edge-on.
From this driving source, the jet extends along a position angle of $\approx$~30$\degr$.
The closest pair of shock emission knots is about 5$\arcsec$ to the north and south of the
driving source. A second pair of emission features $\approx$20$\arcsec$ to the north and south
appears less symmetric, with the southern shock extending further away from the source.

\subsubsection{HOD07 6}
To the west of the center of Fig. 9 and in Fig. 10 (top panels), 
a faint string of four 2.12~$\mu$m S(1) emission knots (labeled 6) and one larger 
emission knot further to the south west
at 6$^h$7$^m$38$\stackrel{s}{.}$7 -6$\degr$11$\arcmin$59$\arcsec$ 
are all roughly aligned at P.A. $\approx$~45$\degr$. 
The position of the central source cannot be determined with any certainty, and the Spitzer
images do not provide additional information in this case.
For lack of any better information, we list this jet by the coordinate of the approximate
center of symmetry of the four northernmost emission knots located at 6$^h$7$^m$41$\stackrel{s}{.}$2, -6$\degr$11$\arcmin$14$\arcsec$, 
implicitly assuming this as
a plausible position of a central source if these knots are, as in many other cases,
symmetric shock fronts in a bipolar jet.

\subsubsection{HOD07 7} 
The position given for this object in Table 1
is that of a faint star at 6$^h$7$^m$41$\stackrel{s}{.}$5, -6$\degr$21$\arcmin$28$\arcsec$
roughly between two bright knots
of S(1) emission in Fig.~7 and Fig.~8 (bottom panels) 
that define a jet axis along P.A. $\approx$~30$\degr$. 
Faint filamentary emission extends back from the shocks
to the immediate vicinity of this star, strongly suggesting
that it is the source of the jet.
On the 2.12~$\mu$m image, the northern jet bends to the east and the southern jet to the west about
8$\arcsec$~-~10$\arcsec$ from the star, suggesting a strongly precessing jet.
The two brightest S(1) emission knots are located near the bend in the jet (in the north) and slightly
beyond it in the south. 
This jet morphology can be understood by assuming that the bent, contiguous jet was formed
by a precessing driving source. The two emission knots lying outside of the contiguous jets
could be the product of a single burst of higher outflow velocity. 
The fact that these higher velocity knots were emitted in the same
projected direction as the strongest bend in the jet may indicate that the
strong precession of the driving source's accretion disk that caused this
bend may also have caused the higher accretion rate that in turn produced the higher
outflow intensity and velocity found in the two brighter knots.

The IRAC 3.6 $\mu$m (not shown here) and 4.5 $\mu$m (Fig.~8, bottom center) images
show the strongest, and spatially extended emission 
associated with the northern of the two S(1) emission regions. 
To the east
of this object, extended emission is seen both on the S(1) image and 
on the IRAC 4.5 $\mu$m and 8.0 $\mu$m images, and this
emission appears to be associated with a filament extending from the Mon~R2 central cluster.  
Whether this filament is physically close to the driving source of the outflow or whether the
outflow interacts with the filament cannot be conclusively determined.
However, the different emission flux levels in the northern and southern lobe could be understood
if the northern jet ran into denser material.

\subsubsection{HOD07 8}
On the 2.12~$\mu$m image in Fig.~9, this jet extends from an unresolved object
at 6$^h$7$^m$43$\stackrel{s}{.}$7, -6$\degr$10$\arcmin$46$\arcsec$ 
to a knot at 6$^h$7$^m$44$\stackrel{s}{.}$4, -6$\degr$10$\arcmin$11$\arcsec$, a
position angle of 16$\degr$ and 
distance of 36$\arcsec$, corresponding to a projected length of 0.14 pc.
The jet axis is outlined by extended 2.12~$\mu$m emission.
The only bright 8.0 $\mu$m object (Fig.~10) associated with this jet is the
southern emission knot.
It is therefore likely that this
8.0 $\mu$m object is the central star driving a jet where predominantly 
only one side is visible. 
The position of this bright mid-infrared source is listed as
the position of the jet in Table~1 and shown in Fig.~9 and Fig.~10.
On the 2.12~$\mu$m image, emission extends about 2.5$\arcsec$ to the 
south of the unresolved central source, suggesting that this is
part of the jet's southern lobe. 

\subsubsection{HOD07 9}
In Fig. 9, 
a well developed bow shock 
is located to the west of a
star at 6$^h$7$^m$44$\stackrel{s}{.}$0, -6$\degr$11$\arcmin$2$\arcsec$.
An emission feature behind the bow shock, near the center of the arc formed by the bow, could be a 
Mach disk associated with the bow shock.
The driving source of this bow shock cannot be determined with certainty from
the data available to us. We are discussing two possible identifications for the
driving source:
The most plausible candidate for the driving source of this bow shock is the star
at 6$^h$7$^m$44$\stackrel{s}{.}$0, -6$\degr$11$\arcmin$02$\arcsec$ that 
is the second brightest 4.5 $\mu$m and 8.0 $\mu$m source in
Fig.~10 and
that shows extended emission on the S(1) image indicating an illuminated outflow
cavity, opening towards the west. 
We use this position 
at 6$^h$7$^m$44$\stackrel{s}{.}$0, -6$\degr$11$\arcmin$2$\arcsec$ 
as the nominal position of this jet in Table~1.

A much fainter, slightly extended object is visible on the UKIRT S(1) image 
at 6$^h$7$^m$43$\stackrel{s}{.}$6, -6$\degr$10$\arcmin$58$\arcsec$ and 
is also
detectable, partly blended with the object described just above, on the Spitzer
4.5~$\mu$m image.
The only reason to consider this object as an alternative candidate for the
driving source of the bow shock is that it is more precisely located on
the symmetry axis of the bow shock.
This object, on the other hand, is also located on
the axis of the jet source HOD07~8, and may just be another shock front of the southern
part of that jet.
The brightest source in Fig.~10 at any of the wavelengths discussed here is a cometary
reflection nebula at 
6$^h$7$^m$43$\stackrel{s}{.}$4, -6$\degr$11$\arcmin$16$\arcsec$
that we list as reflection nebula R5 in Table~2.

\subsubsection{HOD07 10}
Just west of its center Fig.~2 shows 2.12~$\mu$m S(1) emission with the morphology of a jet ending in a
bow shock at 6$^h$7$^m$43$\stackrel{s}{.}$0, -6$\degr$24$\arcmin$20$\arcsec$. 
This object is also shown in the bottom panels of Fig.~3. 
The bow shock is strongly detected in the Spitzer IRAC 4.5 $\mu$m image
and is also indicated in the 8.0$\mu$m image.
It is not clear where the driving source for this jet is located. The jet can be traced to an
extended feature
located at 6$^h$7$^m$44$\stackrel{s}{.}$1, -6$\degr$23$\arcmin$59$\arcsec$
and detected in both the 2.12~$\mu$m and IRAC 4.5~$\mu$m images. 
It is embedded in the emission surrounding the central Mon~R2 cluster.
This may be just another knot in a jet
emerging from deeply inside the molecular cloud, or this may indicate the position of the
driving source. There is no enhanced flux in the 8.0 $\mu$m image at that position.
Nevertheless, we adopt the position of this putative 
driving source as the formal coordinates of this jet called HOD07~10. 
A second shock front, located at 6$^h$7$^m$43$\stackrel{s}{.}$6, -6$\degr$24$\arcmin$11$\arcsec$ 
appears as the brightest of a series of internal shocks along the jet axis, which is oriented
at a position angle of $\approx$~220$\degr$.

\subsubsection{HOD07 11}
The region shown in Fig.~11 contains 2 separate systems of H$_2$ knots shown
in more detail in Fig.~12.
A well-defined bow shock located at 6$^h$7$^m$48$\stackrel{s}{.}$6, -6$\degr$19$\arcmin$03$\arcsec$ and
several fainter knots of 2.12~$\mu$m S(1) emission south-west 
from it suggest a jet with an axis oriented at P.A. $\approx$~23$\degr$. 
The last S(1) emission knot
at 6$^h$7$^m$46$\stackrel{s}{.}$9, -6$\degr$20$\arcmin$6$\arcsec$ 
coincides with a faint Spitzer 4.5 $\mu$m object (Fig.~12, center panel), 
but is most likely
just the 4.5~$\mu$m emission of the same emission knot, and
not the driving source itself. 
The Spitzer 4.5~$\mu$m and 8.0~$\mu$m images in Fig.~12 show a stronger mid-infrared
source at approx. 6$^h$7$^m$46$\stackrel{s}{.}$8, -6$\degr$20$\arcmin$12$\arcsec$ 
just south-west of the position of the last observed S(1) knot in
this string. 
This object is not detected in our
near-infrared S(1) image and therefore must be very deeply embedded. 
We tentatively identify this mid-infrared source as the driving
source of the jet HOD07~11.

\subsubsection{HOD07 12}

The object at 6$^h$7$^m$49$\stackrel{s}{.}$8, -6$\degr$20$\arcmin$43$\arcsec$ in
Fig.~11, HOD07~12,
has the morphology of a faint bipolar nebula seen nearly edge-on, suggesting that it could
be the source of a jet. 
The Spitzer 4.5~$\mu$m image in Fig.~12 shows this object as an unresolved source,
while the 8.0~$\mu$m image does not show the object. 
Two other knots
of S(1) emission lie to the south-east at 6$^h$7$^m$51$\stackrel{s}{.}$6, -6$\degr$20$\arcmin$58$\arcsec$
and at 6$^h$7$^m$54$\stackrel{s}{.}$9, -6$\degr$21$\arcmin$57$\arcsec$, respectively,
suggesting a jet axis at a position angle $\approx$152$\degr$. While there are
a number of faint emission features to the north-east of the bipolar nebula,
e.g., at 6$^h$7$^m$48$\stackrel{s}{.}$8, -6$\degr$20$\arcmin$23$\arcsec$,
no clear association of these features with the jet source can be established.
We tentatively identify the bipolar nebula as the driving source of
HOD07 12 based on its morphology.
A much brighter star, also associated with extended emission at 2.12~$\mu$m and in the
Spitzer bands, and located
at 6$^h$7$^m$49$\stackrel{s}{.}$14, -6$\degr$20$\arcmin$33$\arcsec$, lies also on the
jet axis and is surrounded by some 2.12~$\mu$m emission. 
It is the brightest source in the area on the Spitzer 8.0 $\mu$m image,
indicating that it is a luminous, deeply embedded object. 

\subsubsection{HOD07 13}
The nearly symmetric system of H$_2$ shocks labeled HOD07~13
and located south of the Mon~R2 cluster is the largest jet
found in our survey.
The center of the outflow HOD07~13
is clearly identifiable by an object
on our S(1) image (Figs. 13 and 14) at position
6$^h$7$^m$57$\stackrel{s}{.}$4, -6$\degr$31$\arcmin$6$\arcsec$. 
From this object, the northern jet extends along a position angle
of $\approx$~-13$\degr$.
Unfortunately, the available 4.5~$\mu$m and
8.0~$\mu$m Spitzer images do not extend sufficiently far south to include this object. 
However, all 2.12~$\mu$m features of this object are confirmed by the 
Spitzer 3.6~$\mu$m image (Fig. 13, right panel) that
shows essentially the same outflow shock features. 

A magnified view of the central region of this outflow is shown in Fig.~14.
Immediately north of the central
object, a faint string of emission knots is visible stretching about 15$\arcsec$ to the
north. There is no counterpart to this knotty jet to the south of the central source,
indicating the the jet is inclined with its northern part towards the observer, and that
the southern counterjet is obscured by molecular material surrounding the central source.
However, the IRAC 3.6 $\mu$m image shows a faint outline of a paraboloid outflow cavity wall
to the south. 

More distant S(1) emission shock regions show a remarkable level of symmetry, since they are
not affected by the obscuring material surrounding the central source. 
The first pair of emission regions
are two complex regions, each showing two strands of emission knots, centered roughly at
6$^h$7$^m$56$\stackrel{s}{.}$3, -6$\degr$29$\arcmin$39$\arcsec$ (labeled 1N) 
and 6$^h$7$^m$57$\stackrel{s}{.}$9, -6$\degr$32$\arcmin$23$\arcsec$ (labeled 1S). 
Further out, we find a symmetric pair of fainter shock fronts at
6$^h$7$^m$55$\stackrel{s}{.}$7, -6$\degr$29$\arcmin$4$\arcsec$ (labeled 2N) 
and 6$^h$7$^m$58$\stackrel{s}{.}$3, -6$\degr$32$\arcmin$56$\arcsec$ (labeled 2S). 
The shock fronts farthest removed from the central source show substructure that is
again similar in both the northern and southern jet: A symmetric pair of shock fronts
at 6$^h$7$^m$54$\stackrel{s}{.}$1, -6$\degr$27$\arcmin$33$\arcsec$ (labeled 3N) 
and 6$^h$8$^m$00$\stackrel{s}{.}$2, -6$\degr$34$\arcmin$30$\arcsec$ (labeled 3S). 
Finally, the leading shock fronts are found at 
6$^h$7$^m$53$\stackrel{s}{.}$8, -6$\degr$27$\arcmin$19 (labeled 4N) 
and 6$^h$8$^m$01$\stackrel{s}{.}$0, -6$\degr$34$\arcmin$43 (labeled 4S).
The total projected length of this jet is $\approx$7.5$\arcmin$, corresponding to 1.8~pc.

\subsubsection{HOD07 14}
The compact, bipolar jet HOD07 14~(Fig.~15 and Fig.~16 top panel) 
is characterized by two bright shock fronts,
at 6$^h$7$^m$57$\stackrel{s}{.}$8, -6$\degr$25$\arcmin$26$\arcsec$ and 
6$^h$8$^m$00$\stackrel{s}{.}$0, -6$\degr$25$\arcmin$40$\arcsec$, respectively,
that define a jet axis at position angle $\approx$~114$\degr$.
The Spitzer 4.5~$\mu$m image (Fig.~16, top row, center) confirms the emission seen
at 2.12~$\mu$m, but also shows additional emission, possibly indicative
of a second jet parallel to HOD07~14, but more deeply embedded and therefore not visible
in the near-infrared.
The only indication of the position
of the driving source of HOD07~14 is a flux minimum in the Spitzer 8.0~$\mu$m image,
possibly a dense molecular clump seen in absorption against background
emission. This argument would place the driving source at approximately
6$^h$7$^m$58$\stackrel{s}{.}$9, -6$\degr$25$\arcmin$31$\arcsec$, the position
listed in Table~1 and marked by a circle in Fig.~16.

\subsubsection{HOD07 15}
The UKIRT S(1) image (Fig.~15) shows a nearly symmetric, bent system of shock fronts,
with two bright shocks both to the north and south of the assumed position of the driving source.
In addition, on each side, one fainter, extended shock is visible further away from the
center of symmetry.
The Spitzer 4.5 $\mu$m image (Fig.~16, bottom row, center) shows two additional emission knots in the jet,
closer to the center,
while the Spitzer 8.0 $\mu$m image shows a flux minimum in between those
two strongly obscured 4.5 $\mu$m shocks. We identify this position, approximately
6$^h$8$^m$10$\stackrel{s}{.}$0, -6$\degr$24$\arcmin$47$\arcsec$, 
as the position of the driving source of the jet
HOD07 15 in Table~1 and Fig.~15.

\subsection{Other Bipolar and Cometary Nebulae}
The area near the Mon~R2 cluster contains numerous small patches of
nebulosity, often with bipolar or cometary morphology, that did not
fit our criteria for identification as shocks or jets. We list those objects in Table~2
and indicate them by small square symbols in Figs.~1 and 2. 
The bipolar or cometary shape of many of these objects suggests that these are young,
embedded stars still surrounded by disks and that they have just excavated an outflow
cavity in the surrounding molecular material. 
These objects appear particularly numerous in the area about 2 to 3 arcminutes
south of the main Mon~R2 cluster, and just north of the cluster, as is demonstrated in Figs. 1 and 2.
This finding is consistent with the result by \citet{gut05} that the Class I sources
in Mon~R2, identified by their J, H, and K, and Spitzer IRAC colors, are concentrated in a filamentary
distribution to the south of the cluster center, and to the north and north east of the cluster. 
While there is no strict relationship
between reflection nebulae of bipolar or cometary morphology and SED Class I, the two
criteria cover objects of similar evolutionary status at the trailing edge of their 
accretion phase. 
The main difference between the distribution of reflection nebulae and of the Class I sources
identified by \citet{gut05} is that we find fewer individual reflection nebulae in the region
dominated by the large, ring-like reflection nebula and the filamentary emission that dominates
the center of the cluster. We attribute this difference largely to the difficulty of identifying
small reflection nebulae around individual stars by morphological criteria in this crowded central
region of the Mon~R2 cluster.

By comparison, \citet{gut05} found that the older Class II sources,
identified by J, H, and K-band colors, are more uniformly distributed around the
Mon~R2 cluster center, indicating a dynamically more relaxed population of more developed
young stars.

\section{OTHER INTERESTING OBJECTS IN THE FIELD}

\subsection{HH 866}
The group of Herbig-Haro objects HH~866 found by \citet{wan05} in the L~1646 cloud 
was included in our field (Figs. 1 and 17). 
The HH objects are associated with the IRAS source 06046-0603. The region was
identified by \cite{car00} as a region of enhanced star density in 2MASS data,
and was identified as a potential CO outflow source by \citet{xu01}.
Our UKIRT/WFCAM 2.12 $\mu$m image shows a system of features that can be morphologically
identified as likely H$_2$ shocks. 

The main system of emission knots in the center of Fig.~17 is
extended roughly in north-south direction, and is probably physically associated with the
bow shock at 
6$^h$7$^m$07$\stackrel{s}{.}$5, -6$\degr$04$\arcmin$36$\arcsec$. 
The position of the driving source of this jet cannot be determined with certainty.
We tentatively list the center of symmetry of the brightest H$_2$ emission knots
at 
6$^h$7$^m$07$\stackrel{s}{.}$9, -6$\degr$03$\arcmin$42$\arcsec$ as the
position of the driving source, and name this jet HH 866 Jet W in Table~1. 
The central parts of this north-south jet are coinciding closely with
knots A, B, and C in the H$\alpha$ and [SII] images of \citet{wan05}.
To the east of this position is a large, rather poorly defined bow shock at
6$^h$7$^m$08$\stackrel{s}{.}$1, -6$\degr$03$\arcmin$36$\arcsec$ that appears to
be associated
with more shock emission knots further east of it. A plausible identification of the driving
source is a star embedded in nebulosity at 
6$^h$7$^m$09$\stackrel{s}{.}$8, -6$\degr$03$\arcmin$44$\arcsec$. 
The nebulosity around this star was also noted by \citet{wan05} and the star
is labeled ''1'' in their H$\alpha$ image. Our image resolves this object
into two stars, the eastern and fainter of which is associated with the nebulosity and is the
likely outflow source. We list the
position of this star as HH 866 Jet E in Table~1. 

\subsection{GGD~11}
The bipolar reflection nebula GGD~11 \citep{gyu78} was included in our UKIRT/WFCAM image
and in Fig.~1.
We show a more detailed view of this object in Fig.~18, because our image is, to our knowledge,
the best available near-infrared image of GGD~11. This object has
the typical morphology of a bipolar nebula, with the eastern lobe being
far brighter than the western lobe, due to inclination of the object.
The brightness distribution is fairly smooth,
in contrast to the knotty and filamentary appearance of typical H$_2$ jets.

\subsection{NGC~2182}
The NGC~2182 reflection nebula was included in our UKIRT/WFCAM image
even though it is not included the image section shown in Fig.~1.
It is shown in detail in Fig.~19.
Different from the smooth appearance of this
object on optical photographs, e.~g. on the red plate of the DSS2, and on the K-band 2MASS image, 
the UKIRT 2.12 $\mu$m S(1) image shows two
filamentary regions of enhanced S(1) line emission to the west of the
illuminating star. We cannot determine the excitation mechanism
of these two filaments just from one image and both fluorescent excitation
and shock excitation are possible.

\section{COMPARISON WITH OTHER STAR FORMING REGIONS}

In our 2.12~$\mu$m imaging survey of the Mon~R2 molecular cloud for H$_2$ jets associated
with forming stars, we have found a total of 15 new H$_2$ jets in Mon~R2 and two additional
H$_2$ jets in the L~1646 (HH~866) region. Together with the two
outflows detected in CO by \citet{taf97} and the IRS~3 microjet found by \citet{pre02}
this gives a minimum of 20 active jets and outflows in the larger area of Mon~R2 (including
the L~1646 region), in addition to the
huge, but probably inactive, CO outflow found by \citet{lor81}. 
The spatial distribution of the H$_2$ jets in Mon~R2 along the projected edge
of the large CO outflow, in particular the cluster
of H$_2$ jets 12$\arcmin$ north of the Mon~R2 main cluster, suggests
that this star-formation activity was triggered by the huge fossil CO
outflow in Mon~R2, one of the largest outflows known.

By comparison, in a similar survey of Orion A, \citet{sta02} found 44 jets with
a high degree of certainty, and 29 less certain cases. Considering that their
survey area was larger than ours, and the Orion A cloud is more massive than the
Mon~R2 cloud, this comparison implies a similar specific star formation rate in
both clouds. 
The level of outflow activity in Mon~R2 is also similar to that found
in the two massive star-forming regions
DR21 and W75, where \citet{dav07} found a combined number of approximately 50 H$_2$ jet sources
based on a similar UKIRT WFCAM.

NGC~1333 is one of the most active star-forming regions in the solar vicinity
and \citet{bal96} list 15 optical Herbig-Haro objects, H$_2$ jets, and CO outflows
in this region. This region is closer to the Sun than the other regions discussed
before, and the discovery of jets is therefore easier.
In the Perseus molecular cloud as a whole, of which NGC~1333 is a part, \citet{wal05}
counted a total of 141 optically (H$\alpha$ and [SII]) detectable
Herbig-Haro objects that, at the minimum, belong to 30 individual outflows. 
The smaller Barnard 1 cloud ($\approx$ 1200 M$_\odot$) was found by \citet{wal05b}
to have 8 protostars driving outflows. 
Also, \citet{bal06} found a total of at least 20 outflows in the Chamaeleon I
molecular cloud, mostly by an optical search supported by Spitzer data. Chamaeleon I
is a nearby (165 pc), relatively low-mass (1000 M$_\odot$) cloud.
In light of the different techniques used to search for collimated jets from young stars in these
different regions, we conclude that the number of such jets discovered in Mon~R2 is similar
to that found in the other massive star forming regions.

While it may be argued that the presence of reflection nebulosity,
fluorescently excited H$_2$ emission and PAH emission makes the discovery
of shock-excited H$_2$ emission near the central cluster of Mon~R2 difficult, 
this is also the region most intensely studied in the past, and only the IRS~3
microjet was found there.
By contrast, most of our
newly discovered jets are located in the periphery of the central cluster,
and the longest collimated jets were found farthest away from the central
cluster. 
The complete absence of large
H$_2$ jets near the center of the Mon~R2 cluster cannot be explained by observational
selection effects alone and suggests that H$_2$ jets do not form in this environment. 

This is in agreement with the finding by \citet{dav07} in DR21/W75
who conclude that the H$_2$ jets in their sample are usually not
associated with young molecular cores detectable at sub-mm wavelengths 
or with compact infrared clusters.
They further conclude that clustering may inhibit disc accretion 
and the production of extensive flows. 
A similar conclusion can also be drawn from the extensive survey of the
Orion A cloud for H$_2$ jets by \citet{sta02}. In Orion A, H$_2$ jets
are predominantly found outside the area of the Trapezium cluster and the
Orion nebula.
Also, \citet{hod94} noted in their K' imaging survey of CO outflow sources that
member objects of large young clusters are less likely to show
jets, or localized reflection nebulosity than young objects in smaller groupings.

\section{Conclusions}

Our UKIRT WFCAM imaging survey of the Mon~R2 molecular cloud for H$_2$ jets associated
with forming stars was based on a 2.12~$\mu$m H$_2$ 1--0 S(1) plus continuum image. 
Object identification was based on morphology and confirmation by archival 
Spitzer images. We have found a total of 17 new H$_2$ jets, including the two in
L~1646, as well as 27 small reflection nebulae. 

We conclude that the number of H$_2$ jet sources in Mon~R2 is similar to that found
in other star forming regions of comparable size. We note that the H$_2$ jets appear
to be concentrated in projection against the walls of the outflow cavity created by
the large Mon~R2 CO outflow. This is indicative of a scenario of sequential star formation
in Mon~R2 triggered by the large outflow.

We confirm the finding in other high-mass star forming regions that H$_2$ jets tend to
be found outside of the dense central cluster in such regions. It appears that the 
environment of a dense cluster containing some massive stars prevents the formation of large
H$_2$ jets.

\acknowledgments

The United Kingdom Infrared Telescope is operated by 
the Joint Astronomy Centre on behalf of the 
U.K. Particle Physics and Astronomy Research Council. 

This work is based in part on observations made with the
Spitzer Space Telescope, which is operated by the Jet
Propulsion Laboratory, California Institute of Technology
under a contract with NASA.

This publication makes use of data products from the 
Two Micron All Sky Survey, which is a joint project 
of the University of Massachusetts and the 
Infrared Processing and Analysis Center / California 
Institute of Technology, funded by the 
National Aeronautics and Space Administration 
and the National Science Foundation.

\clearpage
\begin{figure}
\figurenum{1}
\includegraphics[scale=0.8,angle=0]{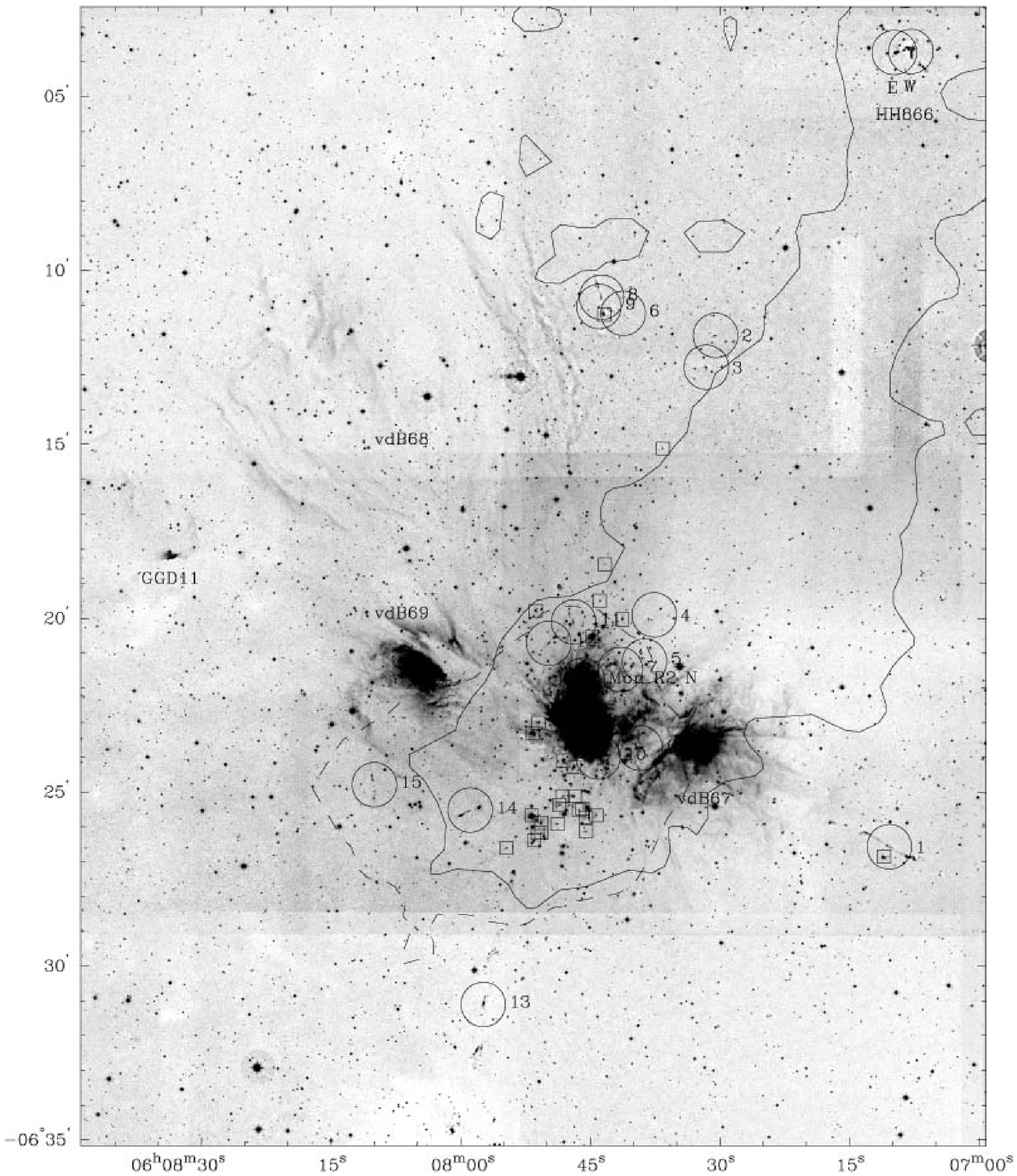}
\caption{
Image of the Mon~R2 star forming region in the 1--0 S(1) emission line of H$_2$ at 2.12 $\mu$m, 
obtained with
the WFCAM at UKIRT. The circles
indicate the positions of the newly found H$_2$ jets. The Herbig-Haro object HH~866 is visible
in the north-west corner of the image. The small reflection nebula near the eastern edge of the
image is GGD~11 \citep{gyu78}. Newly found small reflection nebulae are indicated by squares.
Superposed on this image are lowest contours of the 
blueshifted (-2 to 6 km s${-1}$, solid line) and redshifted ( 14 to 22 km s${-1}$, dashed line) 
CO emission from the map by \citet{wol90}. 
}
\end{figure}

\clearpage
\begin{figure}
\figurenum{2}
\includegraphics[scale=0.8,angle=0]{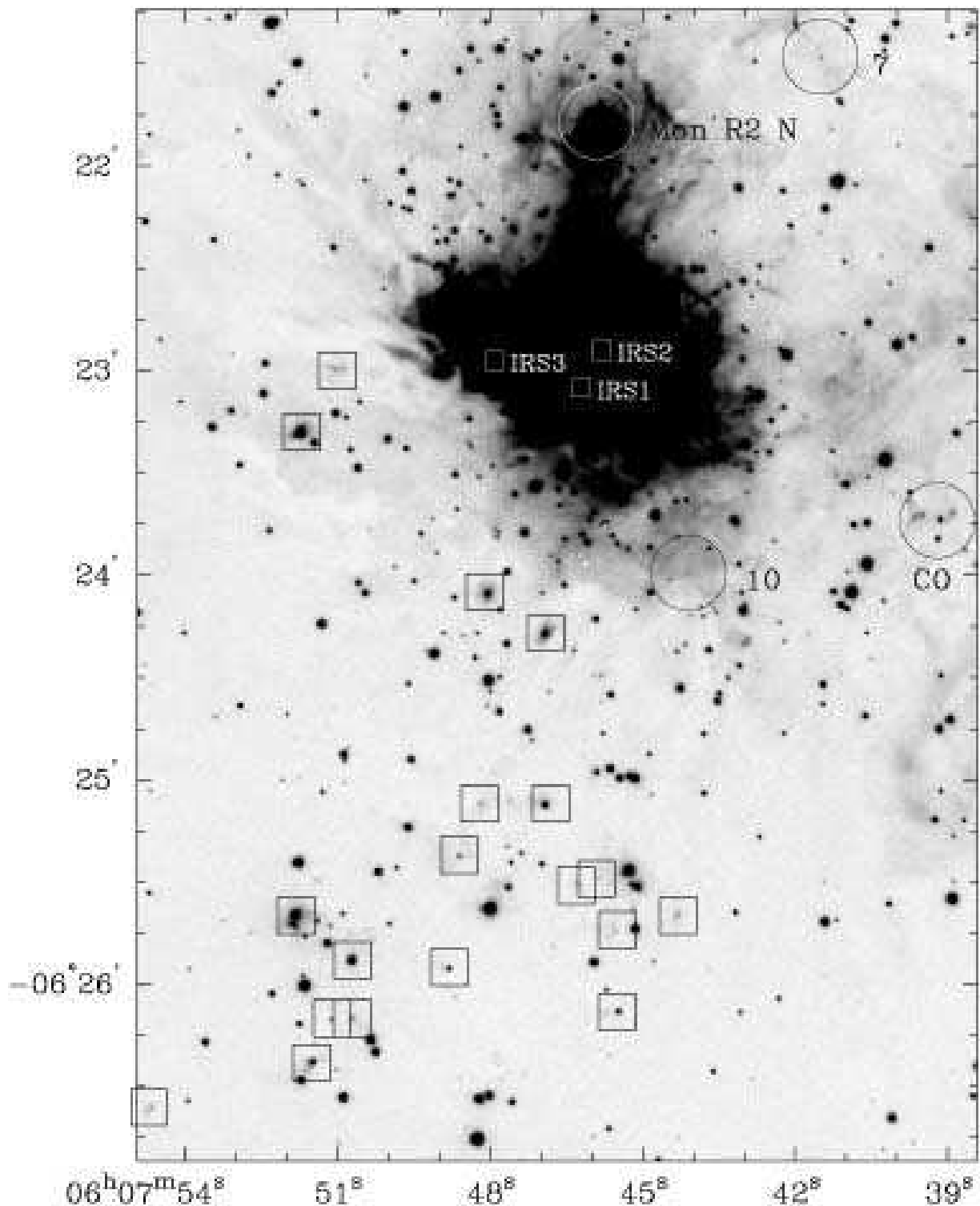}
\caption{
A detailed view of the Mon~R2 center region, based on the same UKIRT/WFCAM
2.12 $\mu$m S(1) line + continuum image as Fig.~1. 
The two CO outflows identified by \citet{taf97}, Mon~R2~N and CO~100,-40, are indicated,
as well as the newly found outflows HOD07 7 and 10.
The positions of the three dominant infrared sources, IRS 1 - 3 \citep{bec76}, are indicated in
the overexposed central region of the Mon~R2 cluster.
The positions of many of the small reflection nebulae from Table~2
are indicated by squares. 
}
\end{figure}

\clearpage
\begin{figure}
\figurenum{3}
\includegraphics[scale=0.8,angle=0]{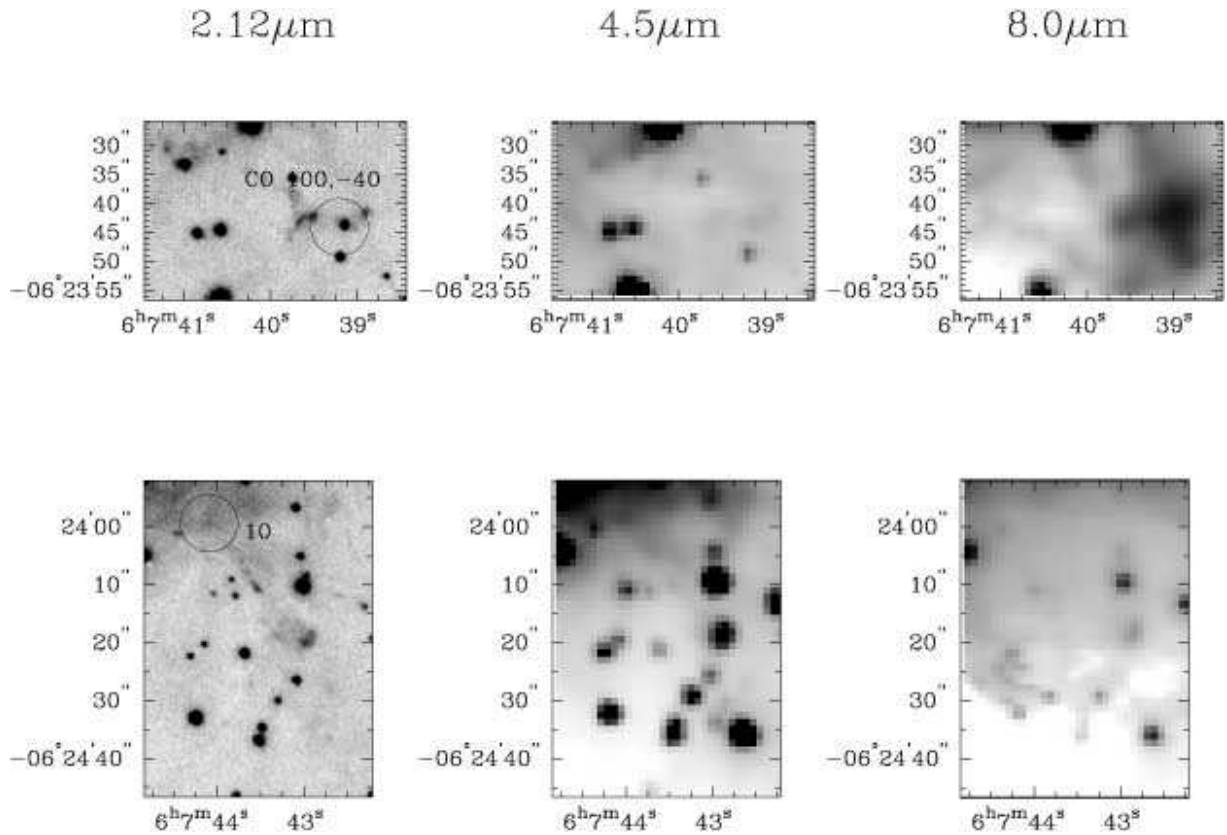}
\caption{
This figure shows detailed views of the two outflows in the 
Mon~R2 center region in the 2.12 $\mu$m S(1) line + continuum, 
Spitzer IRAC 4.5 $\mu$m, and Spitzer IRAC 8.0 $\mu$m bands.
The top panels show the outflow CO 100,-40 \citep{taf97}, the bottom panels the outflow HOD07 10.
}
\end{figure}

\clearpage
\begin{figure}
\figurenum{4}
\includegraphics[scale=0.8,angle=0]{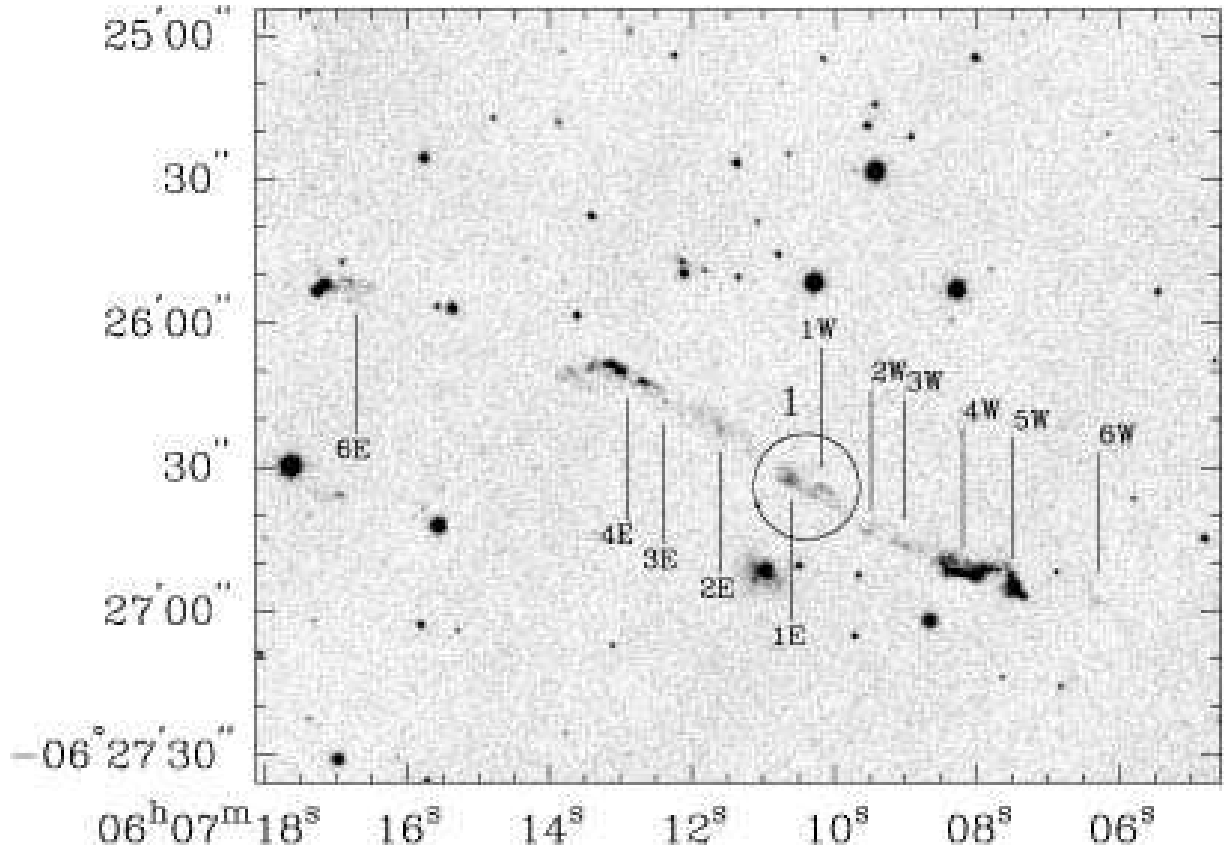}
\caption{
This 2.12 $\mu$m S(1) line + continuum image shows
outflow HOD07~1. The tentative outflow center (marked by a circle) is located at
6$^h$7$^m$10$\stackrel{s}{.}$4, -6$\degr$26$\arcmin$34$\arcsec$ in the flux minimum between the
two patches of nebulosity that define the center of symmetry.
The symmetric sets of shocks in the western and eastern jet are labeled. 
}
\end{figure}

\clearpage
\begin{figure}
\figurenum{5}
\includegraphics[scale=0.8,angle=0]{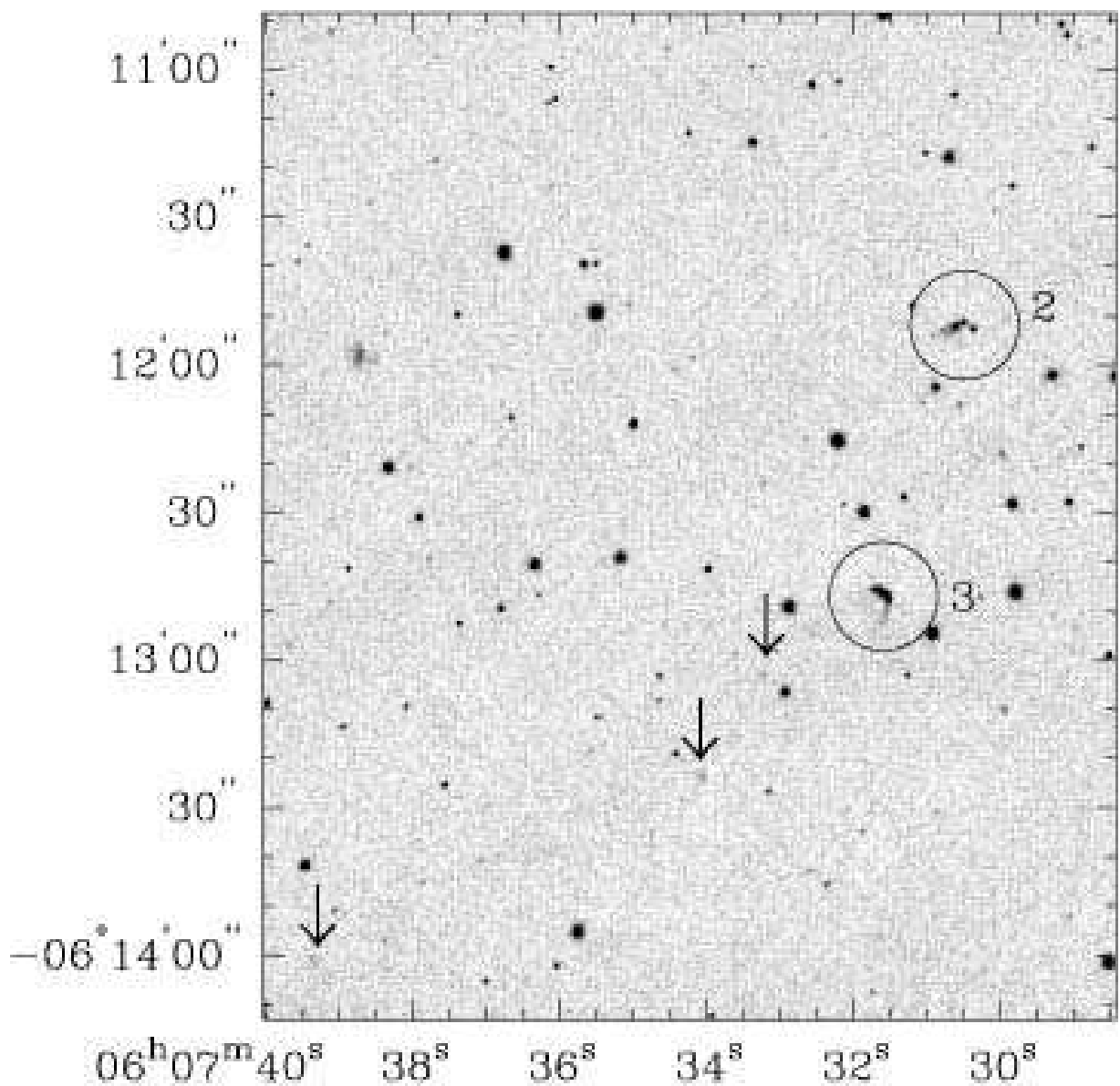}
\caption{
This figure shows the 2.12~$\mu$m image of two 
separate systems of H$_2$ knots:
HOD07~2 at 6$^h$7$^m$30$\stackrel{s}{.}$5, -6$\degr$11$\arcmin$51$\arcsec$
is likely a small jet with several internal shocks.
The bow shock HOD07~3 is located at 6$^h$7$^m$31$\stackrel{s}{.}$6, -6$\degr$12$\arcmin$46$\arcsec$.
Projecting back from the bow shock along its plausible
jet axis, we find a string of faint, slightly extended
knots of emission that are also visible on the
Spitzer 4.5 $\mu$m image. 
These faint emission features, indicated by arrows in Fig.~5,
are located at 6$^h$7$^m$33$\stackrel{s}{.}$2, -6$\degr$13$\arcmin$03$\arcsec$, at 
6$^h$7$^m$34$\stackrel{s}{.}$1, -6$\degr$13$\arcmin$24$\arcsec$, 
and at 6$^h$7$^m$39$\stackrel{s}{.}$3, -6$\degr$14$\arcmin$01$\arcsec$.
Overall, they appear to form a jet with multiple, faint
internal shocks, ending in the bright bow shock.
}
\end{figure}

\clearpage
\begin{figure}
\figurenum{6}
\includegraphics[scale=0.7,angle=0]{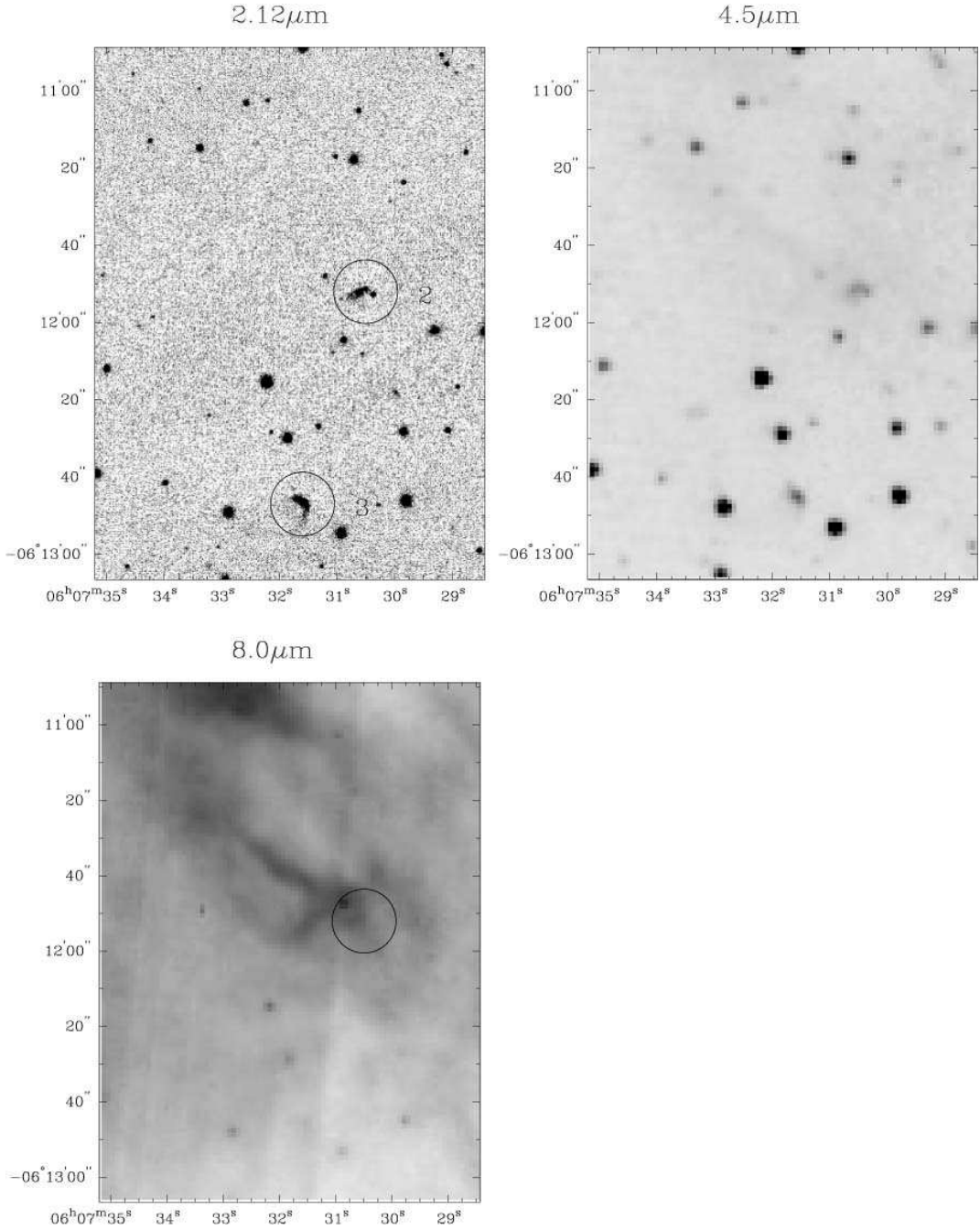}
\caption{
This figure compares the 2.12~$\mu$m, IRAC 4.5~$\mu$m and 8.0~$\mu$m images of two 
separate systems of H$_2$ knots:
HOD07~2 at 6$^h$7$^m$30$\stackrel{s}{.}$5, -6$\degr$11$\arcmin$51$\arcsec$
is a small jet with several internal shocks and 
the position given is that of the most compact knot.
The 8.0 $\mu$m Spitzer
image shows extended nebulosity and a point source that, however, cannot be identified
as the central source of the jet.
The bow shock HOD07~3 is located at 6$^h$7$^m$31$\stackrel{s}{.}$6, -6$\degr$12$\arcmin$46$\arcsec$.
The driving source of this bow shock could not be identified.
}
\end{figure}

\clearpage
\begin{figure}
\figurenum{7}
\includegraphics[scale=0.8,angle=0]{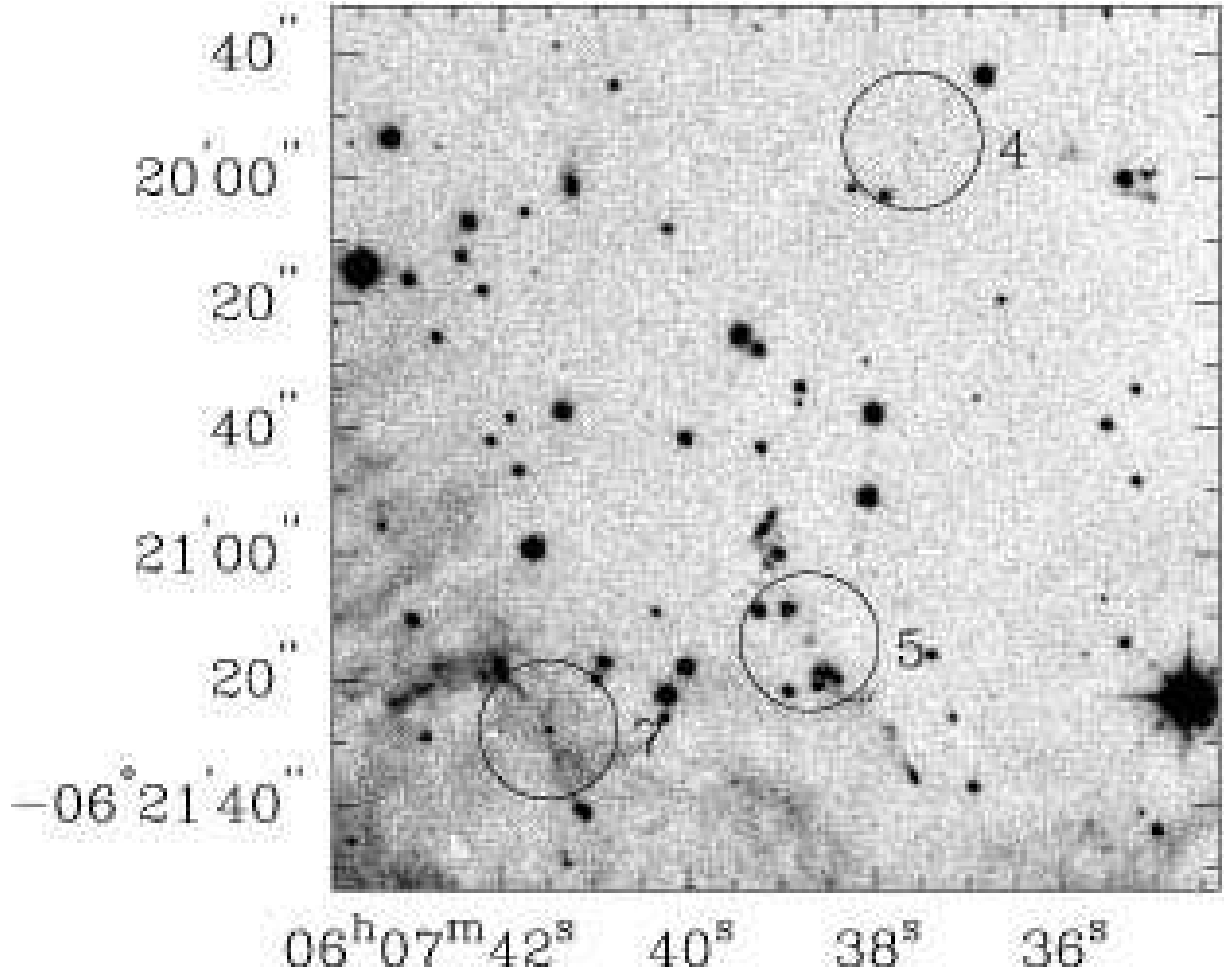}
\caption{
This 2.12 $\mu$m S(1) line + continuum image
contains 3 separate systems of H$_2$ knots:
HOD07~4 at 6$^h$7$^m$35$\stackrel{s}{.}$3, -6$\degr$20$\arcmin$0$\arcsec$, appears as a star with nebulous knots. 
As shown in Fig. 8, this position is most likely not the central source
of this outflow.
HOD07~5 at 6$^h$7$^m$38$\stackrel{s}{.}$7, -6$\degr$21$\arcmin$14$\arcsec$, 
is the likely central star of this bipolar jet.
HOD07~7 at 6$^h$7$^m$41$\stackrel{s}{.}$5, -6$\degr$21$\arcmin$28$\arcsec$, 
is the likely position of the outflow center of that jet.
}
\end{figure}

\clearpage
\begin{figure}
\figurenum{8}
\includegraphics[scale=0.8,angle=0]{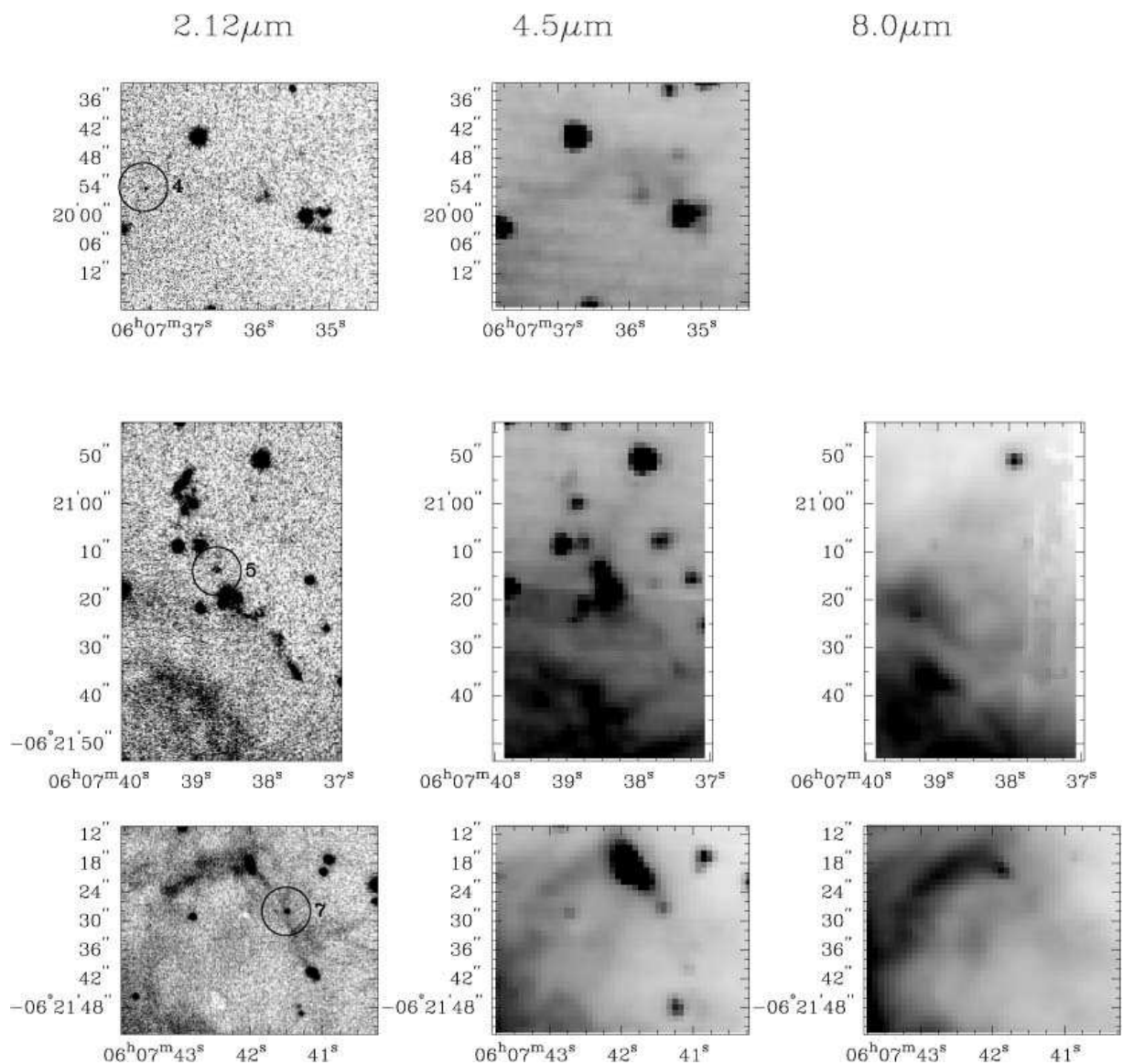}
\caption{
This figure contains the 2.12 $\mu$m S(1), 4.5 $\mu$m and 8.0 $\mu$m images 
of the three systems of H$_2$ knots
found in the area of Fig.~7:
The driving source of HOD07~4 is probably located at 
6$^h$7$^m$37$\stackrel{s}{.}$6, -6$\degr$19$\arcmin$54$\arcsec$,
marked by a very faint star and the apex of the jet-like 
nebulosity of the Spitzer 4.5~$\mu$m image.
HOD07~5 lies at 
6$^h$7$^m$38$\stackrel{s}{.}$7, -6$\degr$21$\arcmin$14$\arcsec$,
identified by a faint source at the approximate center of symmetry
of the bipolar jet.
HOD07~7 at 6$^h$7$^m$41$\stackrel{s}{.}$5, -6$\degr$21$\arcmin$28$\arcsec$, is the likely position of the driving
source of a bipolar jet.
}
\end{figure}

\clearpage
\begin{figure}
\figurenum{9}
\includegraphics[scale=0.8,angle=0]{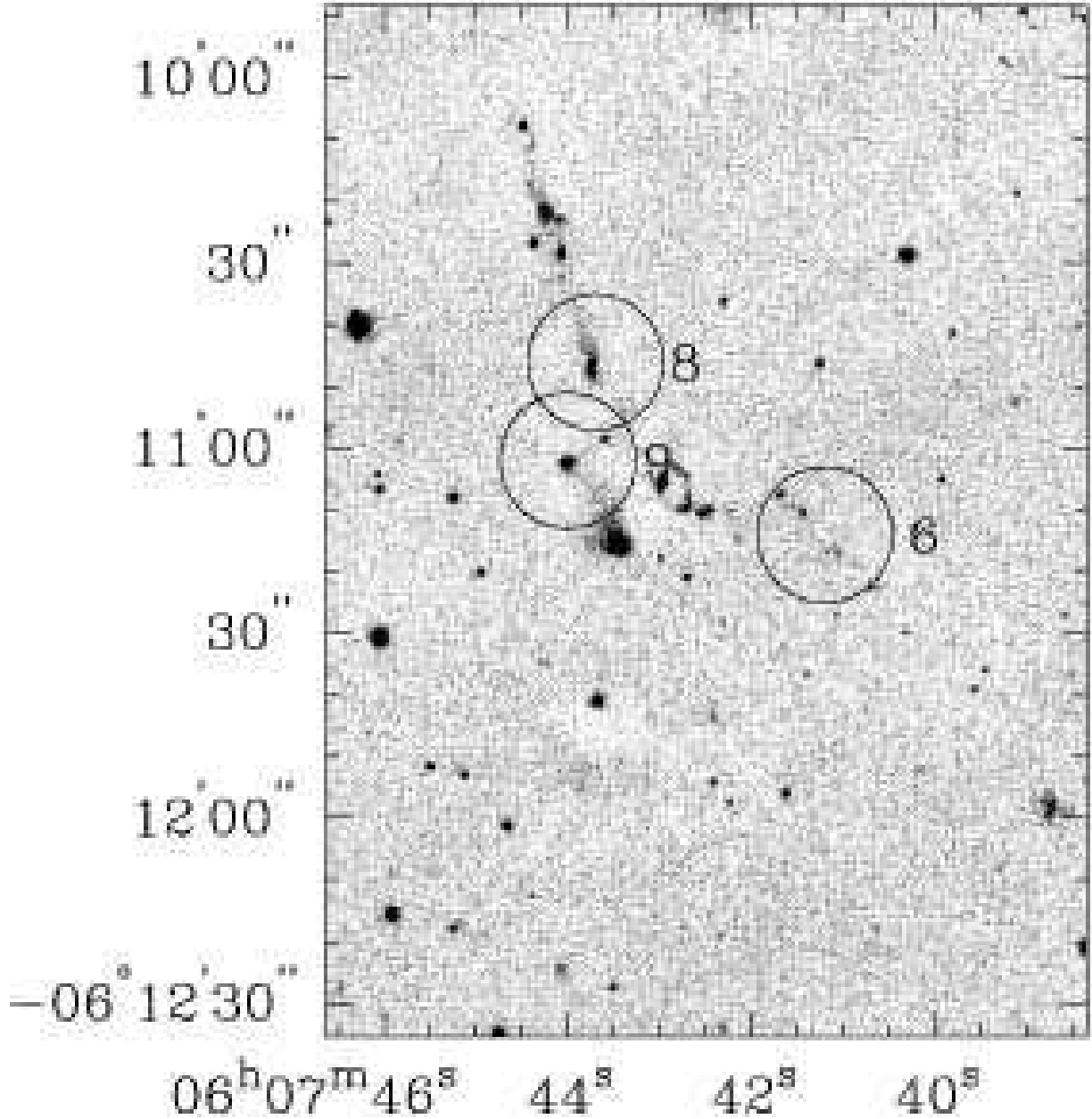}
\caption{
This 2.12 $\mu$m S(1) line + continuum image
contains 3 separate systems of H$_2$ knots:
HOD07~6 at 6$^h$7$^m$41$\stackrel{s}{.}$2, -6$\degr$11$\arcmin$14$\arcsec$ is
the center of symmetry of a string of emission knots.
HOD07~8 at 6$^h$7$^m$43$\stackrel{s}{.}$7, -6$\degr$10$\arcmin$46$\arcsec$ is 
the likely center of a single-sided jet.
HOD07~9 at 6$^h$7$^m$44$\stackrel{s}{.}$0, -6$\degr$11$\arcmin$2$\arcsec$ is 
a star associated with H$_2$ emission and the likely source of a bow shock.
}
\end{figure}

\clearpage
\begin{figure}
\figurenum{10}
\includegraphics[scale=0.8,angle=0]{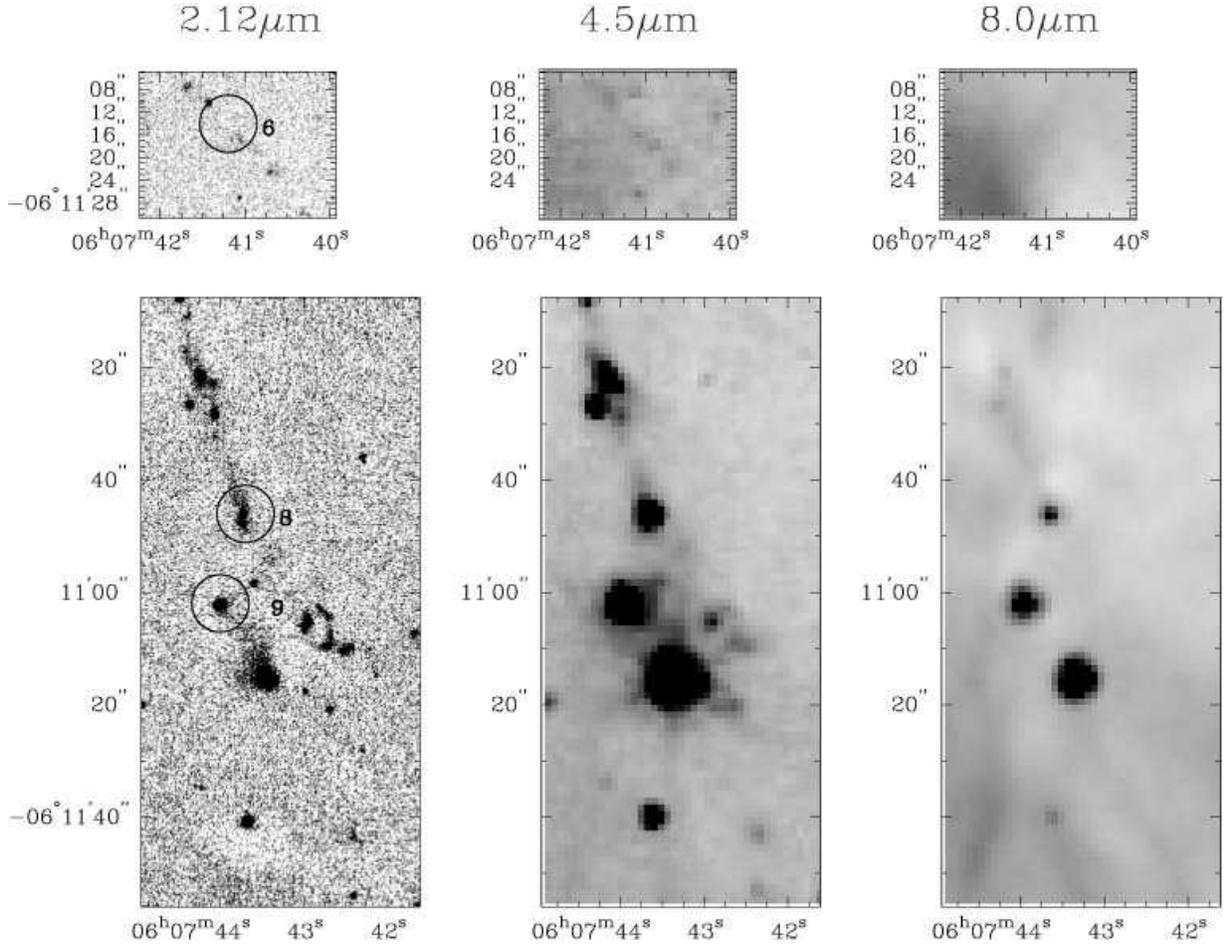}
\caption{
This figure shows the S(1), 4.5 $\mu$m, and 8.0 $\mu$m images of three systems of H$_2$ knots:
In the top row, HOD07~6 at 6$^h$7$^m$41$\stackrel{s}{.}$2, -6$\degr$11$\arcmin$14$\arcsec$ 
is the center of symmetry of a string of emission knots.
In the bottem row, HOD07~8 at 6$^h$7$^m$43$\stackrel{s}{.}$7, -6$\degr$10$\arcmin$46$\arcsec$ is 
the likely center of a single-sided jet.
HOD07~9 at 6$^h$7$^m$44$\stackrel{s}{.}$0, -6$\degr$11$\arcmin$2$\arcsec$ is 
a star associated with H$_2$ emission and the likely source of a bow shock.
}
\end{figure}

\clearpage
\begin{figure}
\figurenum{11}
\includegraphics[scale=0.8,angle=0]{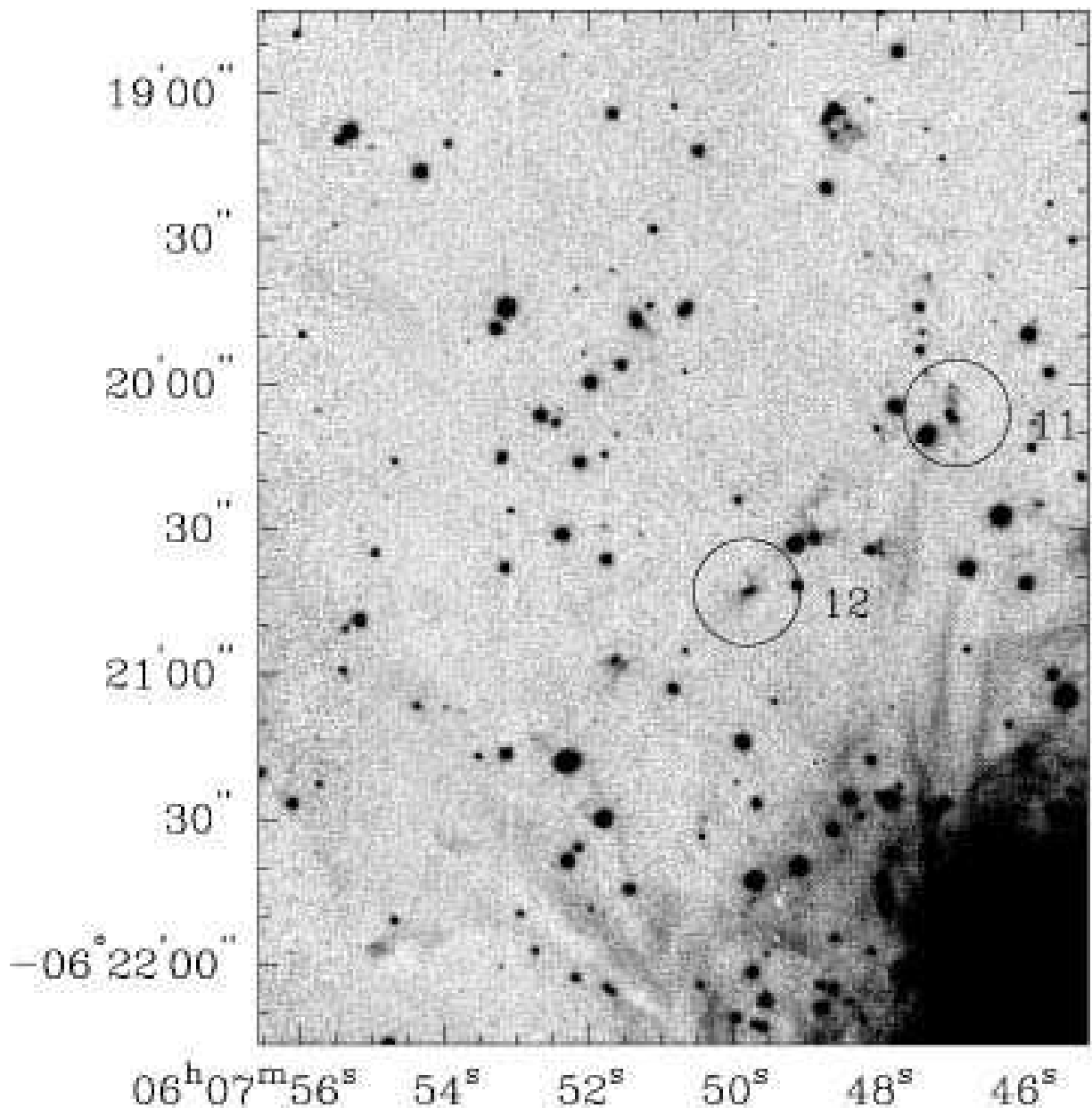}
\caption{
This figure is a 2.12 $\mu$m S(1) line + continuum image
and contains two separate systems of H$_2$ knots:
The circle identifying HOD07~11 at 6$^h$7$^m$46$\stackrel{s}{.}$9, -6$\degr$20$\arcmin$6$\arcsec$
marks the putative position of the 
central source of a jet extending to the north-north-east (P.A. $\approx$23$\degr$) and ending
in a well-defined bow shock located at 6$^h$7$^m$48$\stackrel{s}{.}$6, -6$\degr$19$\arcmin$3$\arcsec$.
Several fainter knots of S(1) emission lie between the center
and the bow shock, completing the morphology of a typical jet.
The object HOD07~12 at 6$^h$7$^m$49$\stackrel{s}{.}$8, -6$\degr$20$\arcmin$43$\arcsec$
has the morphology of a faint bipolar nebula. Two other knots
of S(1) emission lie to the south-east at 6$^h$7$^m$51$\stackrel{s}{.}$6, -6$\degr$20$\arcmin$58$\arcsec$
and at 6$^h$7$^m$54$\stackrel{s}{.}$9, -6$\degr$21$\arcmin$57$\arcsec$, respectively. While there are
a number of faint features to the north-east of the bipolar nebula,
no clear association with this jet source can be established.
}
\end{figure}

\clearpage
\begin{figure}
\figurenum{12}
\includegraphics[scale=0.8,angle=0]{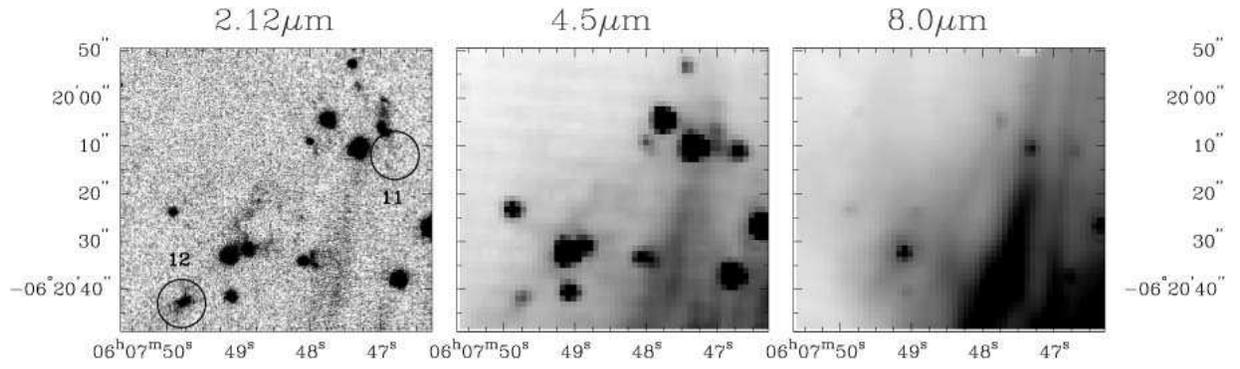}
\caption{
This figure shows images in 2.12 $\mu$m S(1), 4.5 $\mu$m, and 8.0 $\mu$m of 
the central region of the two outflow sources shown in Fig. 11:
HOD07~11 at 6$^h$7$^m$46$\stackrel{s}{.}$8, -6$\degr$20$\arcmin$12$\arcsec$ and
HOD07~12 at 6$^h$7$^m$49$\stackrel{s}{.}$8, -6$\degr$20$\arcmin$43$\arcsec$. 
}
\end{figure}

\clearpage
\begin{figure}
\figurenum{13}
\epsscale{0.7}
\includegraphics[scale=0.8,angle=0]{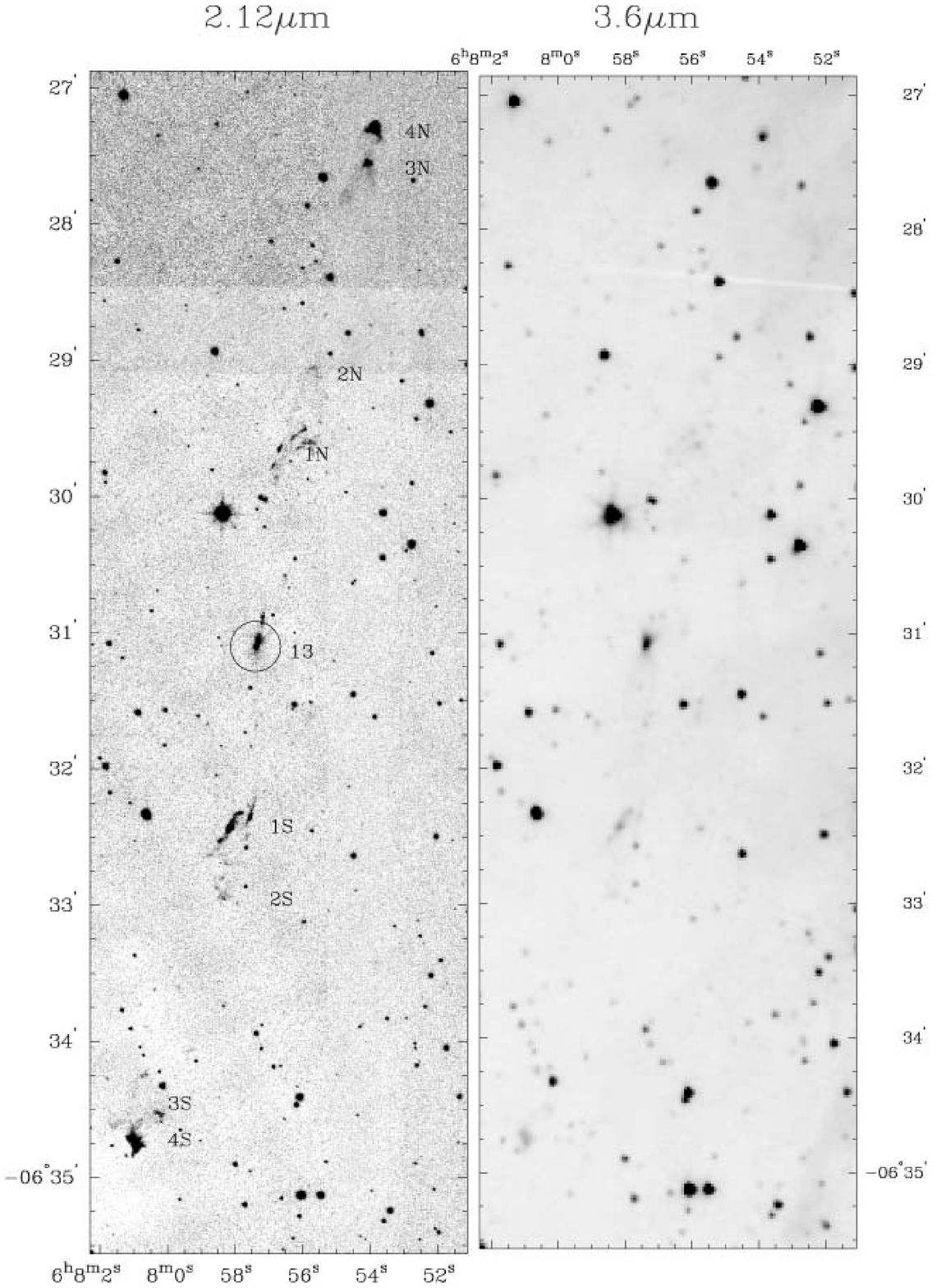}
\caption{
This figure shows the largest of the newly discovered jets:
HOD07~13 with the central source located at 6$^h$7$^m$57$\stackrel{s}{.}$4, -6$\degr$31$\arcmin$6$\arcsec$ 
The jet has a projected length of 7.5$\arcmin$, corresponding
to 1.8 pc.
The left panel shows the UKIRT/WFCAM 2.12 $\mu$m S(1) + continuum image,
the right panel is the Spitzer/IRAC 3.6 $\mu$m image. 
The area of HOD07~13 was not covered by the publicly available
IRAC 4.5 $\mu$m images.
}
\end{figure}

\clearpage
\begin{figure}
\figurenum{14}
\epsscale{0.7}
\includegraphics[scale=0.8,angle=0]{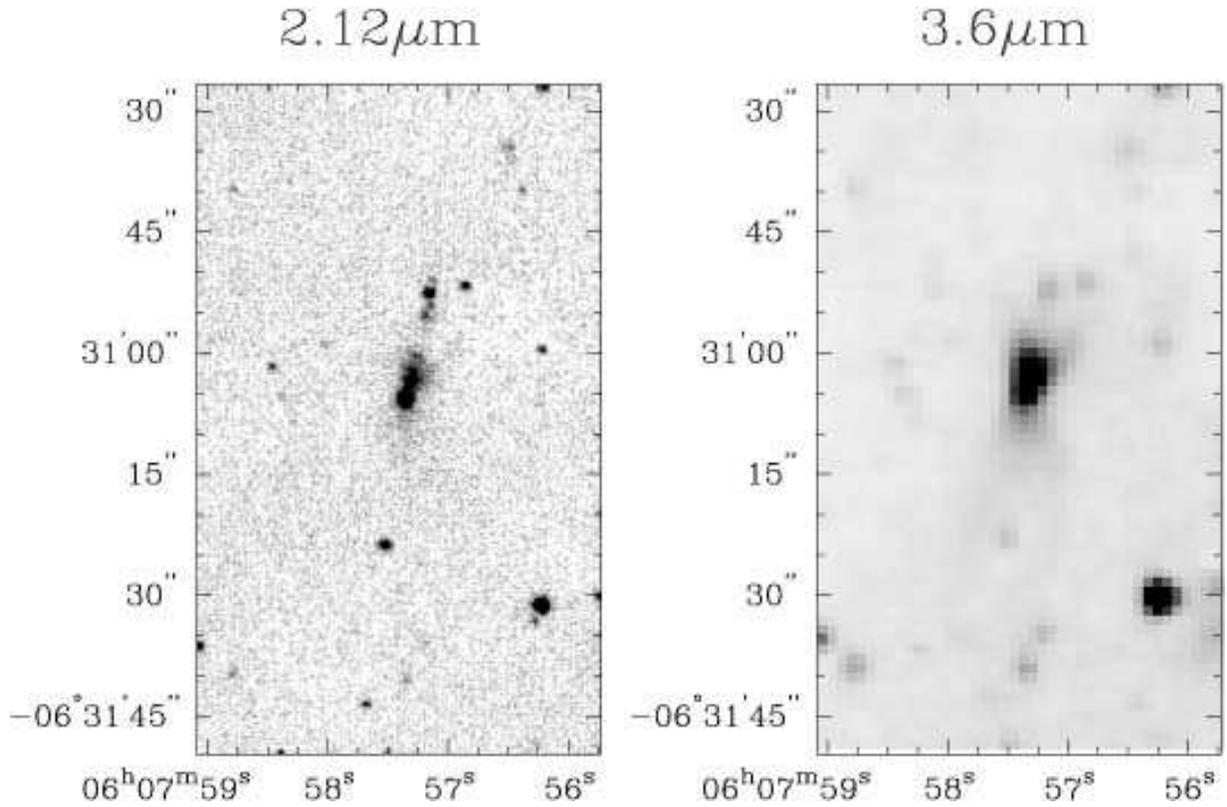}
\caption{
This figure shows the central region of HOD07~13.
The left panel shows the UKIRT/WFCAM 2.12~$\mu$m S(1) + continuum image,
the right panel is the Spitzer/IRAC 3.6~$\mu$m image. 
Both images show a string of shocks north of the position of the central
source of this outflow.
The IRAC 3.6 $\mu$m image shows emission from the southern lobe of the
outflow, indicating the walls of a paraboloidal outflow cavity.
}
\end{figure}

\clearpage
\begin{figure}
\figurenum{15}
\includegraphics[scale=0.8,angle=0]{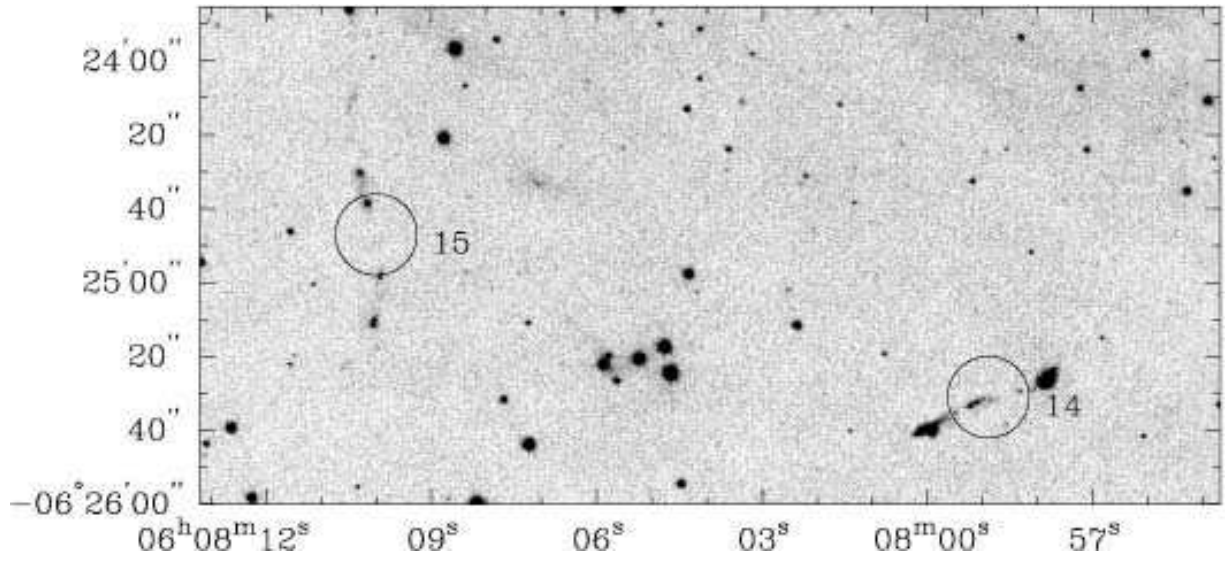}
\caption{
This 2.12 $\mu$m S(1) line + continuum image 
contains two separate systems of H$_2$ knots:
The likely central source of HOD07~14 is located at 6$^h$7$^m$59$\stackrel{s}{.}$2, -6$\degr$25$\arcmin$33$\arcsec$.
The central source of HOD07~15 is located 
near 6$^h$8$^m$10$\stackrel{s}{.}$0, -6$\degr$24$\arcmin$47$\arcsec$. 
}

\end{figure}

\clearpage
\begin{figure}
\figurenum{16}
\includegraphics[scale=0.8,angle=0]{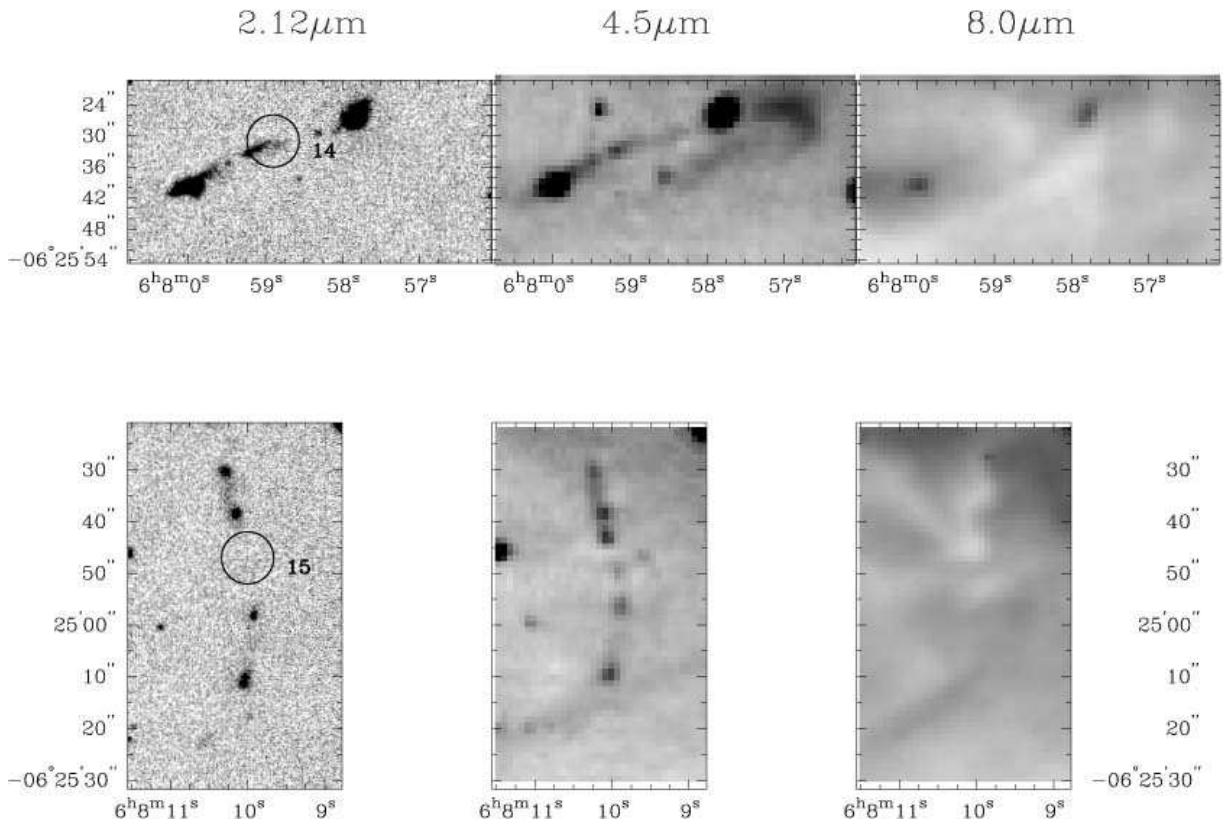}
\caption{
This figure contains the 2.12~$\mu$m, Spitzer 4.5~$\mu$m, and 8.0~$\mu$m images of two separate systems of H$_2$ knots:
The likely central source of HOD07~14 is located at 6$^h$7$^m$59$\stackrel{s}{.}$2, -6$\degr$25$\arcmin$33$\arcsec$.
Based on the symmetry of the emission features, the central source of HOD07~15 located 
near 6$^h$8$^m$10$\stackrel{s}{.}$0, -6$\degr$24$\arcmin$47$\arcsec$. 
}
\end{figure}

\clearpage
\begin{figure}
\figurenum{17}
\includegraphics[scale=0.8,angle=0]{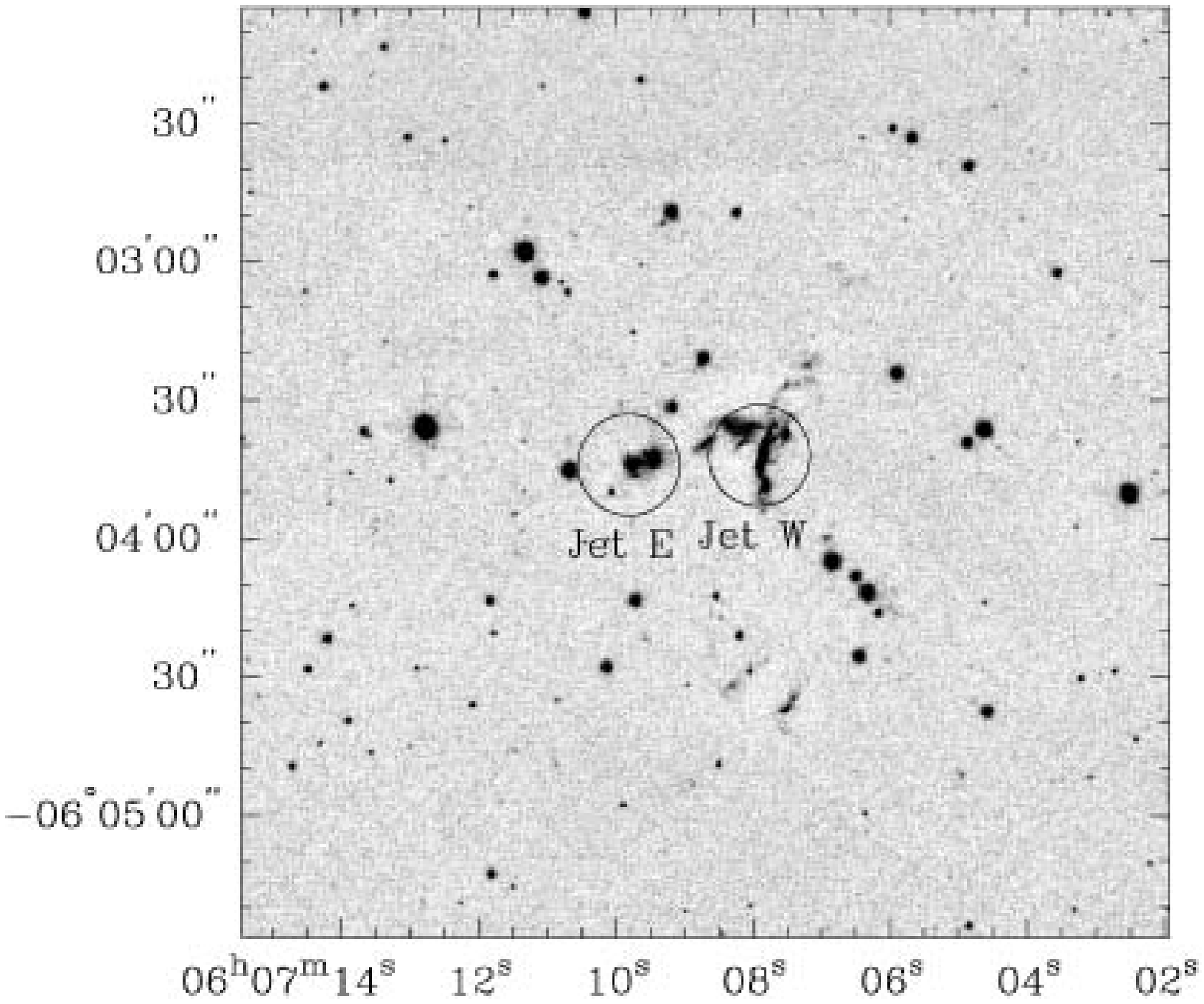}
\caption{
This 2.12 $\mu$m S(1) line + continuum image 
of the nebulosity in L~1646 associated with the optical HH object 866 \citep{wan05}
at 6$^h$7$^m$7$\stackrel{s}{.}$8, -6$\degr$3$\arcmin$49$\arcsec$.
We identify two H$_2$ jets labeled Jet W and Jet E in Fig.~2 and listed in Table~1.
}
\end{figure}

\clearpage
\begin{figure}
\figurenum{18}
\includegraphics[scale=0.8,angle=0]{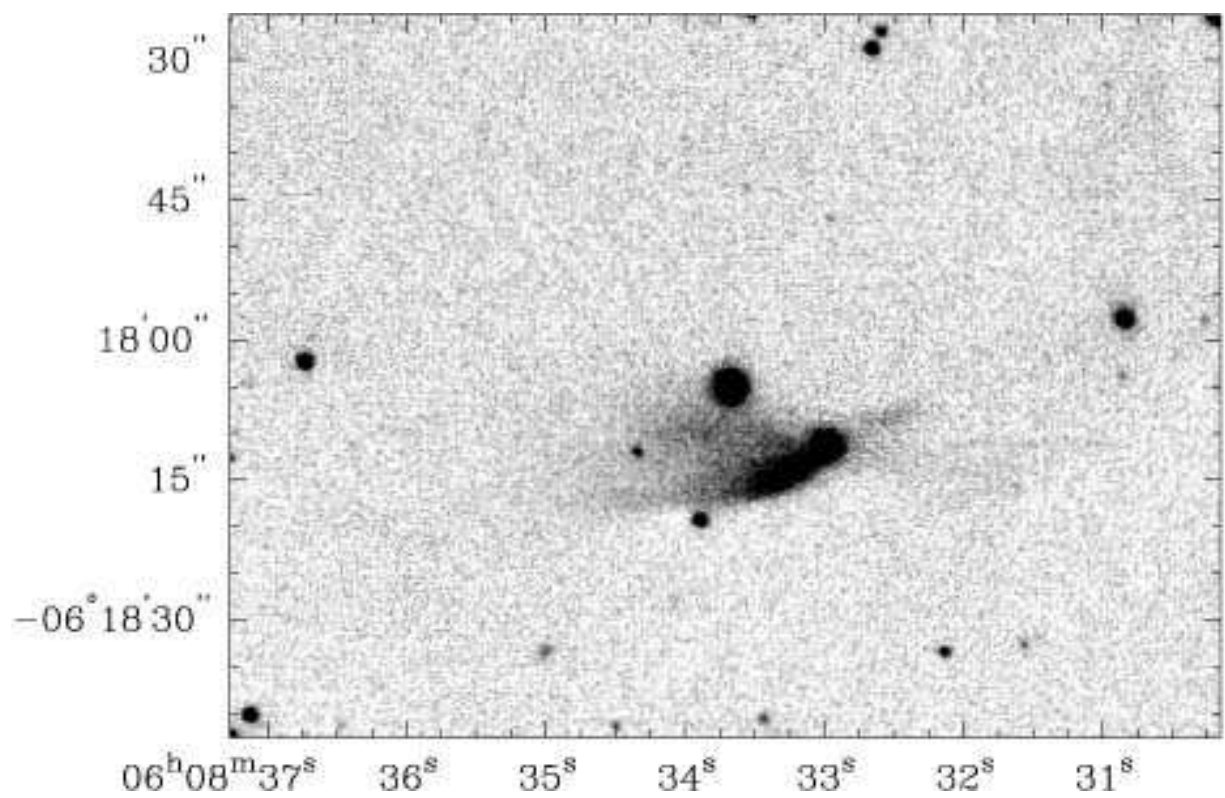}
\caption{
This 2.12 $\mu$m S(1) line + continuum image 
shows the bipolar reflection nebula GGD~11.
}
\end{figure}
\clearpage
\begin{figure}
\figurenum{19}
\includegraphics[scale=0.8,angle=0]{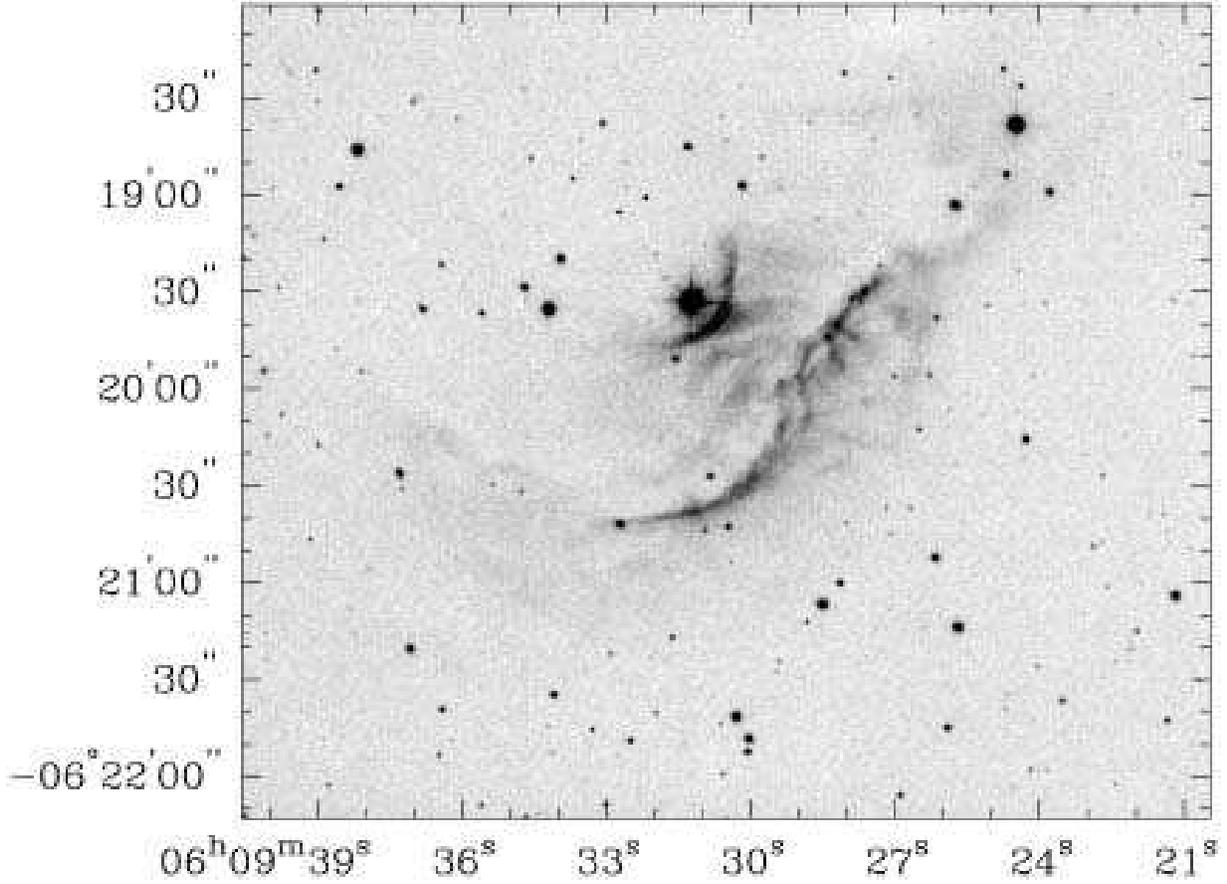}
\caption{
This 2.12 $\mu$m S(1) line + continuum image shows the reflection nebula
NGC 2182, which was recorded in our large WFCAM image but not included
in Fig.~1.  In contrast to optical
and infrared continuum images, this image shows strong, filamentary and
knotty S(1) emission suggesting emission from the shock front of an
expanding shell.
}
\end{figure}

\clearpage
\begin{deluxetable}{ccc}
\tabletypesize{\scriptsize}
\tablecaption{Outflows in Mon R2}
\tablewidth{0pt}
\tablehead{
\colhead{RA (J2000.0)} & \colhead{Dec (J2000.0)} & \colhead{Name}
}
\startdata
6$^h$7$^m$07$\stackrel{s}{.}$9 & -6$\degr$03$\arcmin$42$\arcsec$ & HH 866 Jet W\\
6$^h$7$^m$09$\stackrel{s}{.}$8 & -6$\degr$03$\arcmin$44$\arcsec$ & HH 866 Jet E\\
6$^h$7$^m$10$\stackrel{s}{.}$4 & -6$\degr$26$\arcmin$34$\arcsec$ & HOD07 1\\
6$^h$7$^m$30$\stackrel{s}{.}$5 & -6$\degr$11$\arcmin$51$\arcsec$ & HOD07 2\\
6$^h$7$^m$31$\stackrel{s}{.}$6 & -6$\degr$12$\arcmin$47$\arcsec$ & HOD07 3\\
6$^h$7$^m$37$\stackrel{s}{.}$6 & -6$\degr$19$\arcmin$54$\arcsec$ & HOD07 4\\
6$^h$7$^m$38$\stackrel{s}{.}$7 & -6$\degr$21$\arcmin$14$\arcsec$ & HOD07 5\\
6$^h$7$^m$39$\stackrel{s}{.}$2 & -6$\degr$23$\arcmin$44$\arcsec$ & TBWW97 CO -100,-40\\
6$^h$7$^m$41$\stackrel{s}{.}$2 & -6$\degr$11$\arcmin$14$\arcsec$ & HOD07 6\\
6$^h$7$^m$41$\stackrel{s}{.}$5 & -6$\degr$21$\arcmin$28$\arcsec$ & HOD07 7\\
6$^h$7$^m$43$\stackrel{s}{.}$7 & -6$\degr$10$\arcmin$46$\arcsec$ & HOD07 8\\
6$^h$7$^m$44$\stackrel{s}{.}$0 & -6$\degr$11$\arcmin$02$\arcsec$ & HOD07 9\\
6$^h$7$^m$44$\stackrel{s}{.}$1 & -6$\degr$23$\arcmin$59$\arcsec$ & HOD07 10\\
6$^h$7$^m$45$\stackrel{s}{.}$9 & -6$\degr$21$\arcmin$47$\arcsec$ & TBWW97 Mon~R2-N\\
6$^h$7$^m$46$\stackrel{s}{.}$8 & -6$\degr$20$\arcmin$12$\arcsec$ & HOD07 11\\
6$^h$7$^m$49$\stackrel{s}{.}$8 & -6$\degr$20$\arcmin$43$\arcsec$ & HOD07 12\\
6$^h$7$^m$57$\stackrel{s}{.}$4 & -6$\degr$31$\arcmin$06$\arcsec$ & HOD07 13\\
6$^h$7$^m$58$\stackrel{s}{.}$9 & -6$\degr$25$\arcmin$31$\arcsec$ & HOD07 14\\
6$^h$8$^m$10$\stackrel{s}{.}$0 & -6$\degr$24$\arcmin$47$\arcsec$ & HOD07 15\\
\enddata
\end{deluxetable}

\clearpage
\begin{deluxetable}{cccc}
\tabletypesize{\scriptsize}
\tablecaption{Reflection Nebulae in Mon R2}
\tablewidth{0pt}
\tablehead{
\colhead{RA (J2000.0)} & \colhead{Dec (J2000.0)} & \colhead{Name} & \colhead{Morphology}
}
\startdata
6$^h$7$^m$11$\stackrel{s}{.}$0 & -6$\degr$26$\arcmin$51$\arcsec$ & HOD07 R1 & reflection neb.\\
6$^h$7$^m$36$\stackrel{s}{.}$6 & -6$\degr$15$\arcmin$07$\arcsec$ & HOD07 R2 & bipolar neb.\\
6$^h$7$^m$41$\stackrel{s}{.}$3 & -6$\degr$20$\arcmin$02$\arcsec$ & HOD07 R3 & cometary neb.\\
6$^h$7$^m$43$\stackrel{s}{.}$3 & -6$\degr$18$\arcmin$27$\arcsec$ & HOD07 R4 & reflection neb.\\
6$^h$7$^m$43$\stackrel{s}{.}$4 & -6$\degr$11$\arcmin$16$\arcsec$ & HOD07 R5 & cometary neb.\\
6$^h$7$^m$43$\stackrel{s}{.}$9 & -6$\degr$19$\arcmin$30$\arcsec$ & HOD07 R6 & cometary neb. jet ?\\
6$^h$7$^m$44$\stackrel{s}{.}$3 & -6$\degr$25$\arcmin$40$\arcsec$ & HOD07 R7 & cometary neb.\\
6$^h$7$^m$44$\stackrel{s}{.}$8 & -6$\degr$20$\arcmin$31$\arcsec$ & HOD07 R8 & reflection neb.\\
6$^h$7$^m$45$\stackrel{s}{.}$5 & -6$\degr$25$\arcmin$44$\arcsec$ & HOD07 R9 & reflection neb.\\
6$^h$7$^m$45$\stackrel{s}{.}$5 & -6$\degr$26$\arcmin$08$\arcsec$ & HOD07 R10 & reflection neb.\\
6$^h$7$^m$45$\stackrel{s}{.}$9 & -6$\degr$25$\arcmin$29$\arcsec$ & HOD07 R11 & cometary neb.\\
6$^h$7$^m$46$\stackrel{s}{.}$3 & -6$\degr$25$\arcmin$31$\arcsec$ & HOD07 R12 & reflection nebula\\
6$^h$7$^m$46$\stackrel{s}{.}$8 & -6$\degr$25$\arcmin$07$\arcsec$ & HOD07 R13 & cometary neb.\\
6$^h$7$^m$46$\stackrel{s}{.}$9 & -6$\degr$24$\arcmin$17$\arcsec$ & HOD07 R14 & bipolar neb.\\
6$^h$7$^m$48$\stackrel{s}{.}$1 & -6$\degr$24$\arcmin$05$\arcsec$ & HOD07 R15 & reflection neb.\\
6$^h$7$^m$48$\stackrel{s}{.}$2 & -6$\degr$25$\arcmin$07$\arcsec$ & HOD07 R16 & reflection neb.\\
6$^h$7$^m$48$\stackrel{s}{.}$6 & -6$\degr$25$\arcmin$22$\arcsec$ & HOD07 R17 & bipolar neb.\\
6$^h$7$^m$48$\stackrel{s}{.}$8 & -6$\degr$25$\arcmin$55$\arcsec$ & HOD07 R18 & cometary neb.\\
6$^h$7$^m$50$\stackrel{s}{.}$7 & -6$\degr$25$\arcmin$53$\arcsec$ & HOD07 R19 & reflection neb.\\
6$^h$7$^m$50$\stackrel{s}{.}$7 & -6$\degr$26$\arcmin$10$\arcsec$ & HOD07 R20 & reflection neb.\\
6$^h$7$^m$51$\stackrel{s}{.}$0 & -6$\degr$23$\arcmin$00$\arcsec$ & HOD07 R21 & bipolar neb.\\
6$^h$7$^m$51$\stackrel{s}{.}$1 & -6$\degr$26$\arcmin$10$\arcsec$ & HOD07 R22 & reflection neb.\\
6$^h$7$^m$51$\stackrel{s}{.}$3 & -6$\degr$19$\arcmin$47$\arcsec$ & HOD07 R23 & bipolar neb.\\
6$^h$7$^m$51$\stackrel{s}{.}$5 & -6$\degr$26$\arcmin$23$\arcsec$ & HOD07 R24 & reflection neb.\\
6$^h$7$^m$51$\stackrel{s}{.}$7 & -6$\degr$23$\arcmin$18$\arcsec$ & HOD07 R25 & reflection neb.\\
6$^h$7$^m$51$\stackrel{s}{.}$8 & -6$\degr$25$\arcmin$40$\arcsec$ & HOD07 R26 & reflection neb.\\
6$^h$7$^m$54$\stackrel{s}{.}$7 & -6$\degr$26$\arcmin$36$\arcsec$ & HOD07 R27 & reflection neb.\\
\enddata
\end{deluxetable}

\end{document}